\documentclass[twocolumn,english,aps,prb]{revtex4-1}
\usepackage{color}
\usepackage{amsmath}
\usepackage{graphicx}
\usepackage{amssymb}

\makeatletter
\@ifundefined{textcolor}{}
{%
 \definecolor{BLACK}{gray}{0}
 \definecolor{WHITE}{gray}{1}
 \definecolor{RED}{rgb}{1,0,0}
 \definecolor{GREEN}{rgb}{0,1,0}
 \definecolor{BLUE}{rgb}{0,0,1}
 \definecolor{CYAN}{cmyk}{1,0,0,0}
 \definecolor{MAGENTA}{cmyk}{0,1,0,0}
 \definecolor{YELLOW}{cmyk}{0,0,1,0}
 }

\@ifundefined{definecolor}
 {\usepackage{color}}{}
\@ifundefined{definecolor}{\@ifundefined{definecolor}
 {\usepackage{color}}{}
}{}

\newcommand{\sgn}{\operatorname{sgn}}

\newcommand{\ba}{\begin{eqnarray*}}
\newcommand{\ea}{\end{eqnarray*}}
\newcommand{\baa}{\begin{eqnarray}}
\newcommand{\eaa}{\end{eqnarray}}
\newcommand{\bea}{\begin{eqnarray}}
\newcommand{\eea}{\end{eqnarray}}
\newcommand{\be}{\begin{equation}}
\newcommand{\ee}{\end{equation}}

\newcommand{\sm}{SmB$_6$}

\newcommand{\Ek}{E_{\bk}}
\newcommand{\Vrec}{V_{\rm rec}}
\newcommand{\Emerge}{E_{\rm merge}}
\newcommand{\Ecross}{E_{\rm cross}}
\newcommand{\C}{\mathcal{C}}    % Chern numbers
\newcommand{\aaa}{a_0}           % lattice constant

\newcommand{\bk}{\mathbf{k}}
\newcommand{\bK}{\mathbf{K}}

\newcommand{\bq}{\mathbf{q}}
\newcommand{\bQ}{\mathbf{Q}}

\DeclareMathOperator{\ii}{i}

\makeatother

\usepackage{babel}

\begin{document}

\title{
%Surface reconstruction in a tight-binding model for  {\sm}
Surface reconstruction in a tight-binding model\\for the topological Kondo insulator {\sm}
}

\author{Pier Paolo Baruselli}
\author{Matthias Vojta}
\affiliation{Institut f\"ur Theoretische Physik, Technische Universit\"at Dresden, 01062 Dresden, Germany}

%%%%%%%%%%%%%%%%%%%%%%%%%%%%%%%%%%%%%%%%%%%%%%%%%%%%%%%%%%%%%%%%%%%%%%%

\begin{abstract}
For the strongly correlated topological insulator {\sm} we discuss the influence of a $2\times 1$ reconstruction of the (001) surface on the topological surface states. Depending on microscopic details, the reconstruction can be a weak or a strong perturbation to the electronic states. While the former leads to a weak backfolding of surface bands only, the latter can modify the surface-state dispersion and lead to a Lifshitz transition. We analyze the quasiparticle interference signal: while this tends to be weak in models for {\sm} in the absence of surface reconstruction, we find that the $2\times 1$ reconstruction can induce novel peaks. We discuss experimental implications.
%which are however likely to be too weak to be observed on experiments.
\end{abstract}

\date{\today}


\maketitle
%%%%%%%%%%%%%%%%%%%%%%%%%%%%%%%%%%%%%%%%%%%%%%%%%%%%%%%%%%%%%%%%%%%%%%%

\section{Introduction}

The material {\sm} has attracted significant attention during the last years, as it has been proposed\cite{takimoto,lu_smb6_gutz, tki_cubic} to realize a three-dimensional (3D) topological Kondo insulator (TKI). In this fascinating class of materials, a topologically non-trivial bandstructure emerges at low energies and temperatures as a result of Kondo screening of strongly correlated $f$ electrons.\cite{tki1} Similar to weakly correlated topological insulators (TIs), the surfaces of 3D TKIs display topological surface states with Dirac dispersion and spin-momentum locking.

Recent theory\cite{smb6_tci,smb6_tci_io,smb6_tci_legner} suggested {\sm} to be also a topological crystalline insulator (TCI), its bandstructure being characterized by three non-zero mirror Chern numbers, $\C^+_{k_z=0}$, $\C^+_{k_z=\pi}$, $\C^+_{k_x=k_y}$, defined on planes in the 3D Brillouin zone (BZ) which are invariant under mirror operations.
%Provided the topological character of {\sm},
This allows to further classify topological surface states according to their mirror-symmetry eigenvalues.

On the experimental front, a number of results obtained on {\sm}, e.g., from transport studies,\cite{wolgast_smb6,fisk_smb6_topss, smb6_junction_prx} quantum oscillation measurements,\cite{lixiang_smb6} angle-resolved photoemission spectroscopy (ARPES),\cite{neupane_smb6,mesot_smb6, smb6_arpes_feng,smb6_arpes_reinert,smb6_past_allen} and spin-resolved ARPES \cite{smb6_arpes_mesot_spin}
%and scanning tunneling spectroscopy (STS)\cite{pnas_smb6_stm,hoffman_smb6}
appear consistent with the presence of Dirac-like surface states.
However, doubts have been raised about the proper interpretation of ARPES data.\cite{sawatzky_smb6,smb6_prx_arpes,smb6_trivial}

An important class of experiments, typically used to verify the topological nature of TI surface states, employ scanning tunneling spectroscopy (STS): the surface-state quasiparticle interference (QPI) signal allows one to deduce characteristic wavevectors for spin-conserving elastic scattering due to defects and thus directly probes the spin-momentum locking of surface electrons. Unfortunately, conclusive QPI results on {\sm} are lacking to date.
In fact, surface-sensitive probes such as STS face the problem that {\sm} surfaces are hard to characterize experimentally: Non-reconstructed (001) surfaces are polar, such that surfaces typically reconstruct. A frequent case is the $2\times 1$ reconstructed (001) surface, as observed in large areas in STS experiments \cite{pnas_smb6_stm,hoffman_smb6}. Such a surface is non-polar, rendering it more favorable for the observation of topologically non-trivial surface states.
This implies, however, that a proper theoretical modelling must take into account the surface reconstruction, i.e., the fact that one half of the top Sm rows are missing -- this has not been done in the theory literature to date.

It is the purpose of this paper to fill this gap: We shall study the effects of surface reconstruction of the electronic states of {\sm} on the level of tight-binding models. In particular, we shall focus on the fate of the topological surface states and on signatures in STS and QPI experiments.
Our main results can be summarized as follows. Not unexpectedly, the importance of the surface reconstruction depends strongly on the effective strength of the electronic reconstruction potential. The latter depends on various microscopic details which cannot be reliably extracted from our simplified modelling, such that we rely on assumptions here.
In the case that the reconstruction potential is weak, its effects on QPI are minor despite the backfolding of surface bands, i.e., the effect of reconstruction can -- for most purposes -- be ignored.
If, however, the reconstruction potential is sufficiently strong, a Lifshitz transition -- a change in the Fermi-surface topology as function of the Fermi energy -- of the surface states is expected. Its general properties can be described by appealing to the presence of a mirror plane which prevents band hybridization along certain high-symmetry directions. This is in analogy to other TCI materials\cite{pbsnse_tci,pbsnte_tci,snte_tci} (where a similar scenario applies to non-reconstructed surfaces). Importantly, such $2\times 1$ reconstruction can entirely modify the tunneling signal, and new QPI peaks of unidirectional character may appear.

The remainder of the paper is organized as follows.
We will start with an introductory discussion to {\sm} and its surfaces in Section~\ref{sec_gen}. Section \ref{sec_rec} will then qualitatively discuss the effect of surface reconstruction on the surface states. Concrete numerical results for bandstructures as well as tunneling and QPI spectra will be given in Section \ref{sec_numerics}. Section \ref{sec_concl} summarizes the results of our work.

%%%%%%%%%%%%%%%%%%%%%%%%%%%%%%%%%%%%%%%%%%%%%%%%%%%%%%%%%%%%%%%%%%%%%%%%%
%%%%%%%%%%%%%%%%%%%%%%%%%%%%%%%%%%%%%%%%%%%%%%%%%%%%%%%%%%%%%%%%%%%%%%%%%
%%%%%%%%%%%%%%%%%%%%%%%%%%%%%%%%%%%%%%%%%%%%%%%%%%%%%%%%%%%%%%%%%%%%%%%%%

\section{S\MakeLowercase{m}B$_6$: General remarks}
\label{sec_gen}

To set the stage, we summarize key aspects of {\sm} surfaces and the tight-binding modeling of both bulk and surface electronic states, and we also highlight general aspects of surface reconstruction.

\subsection{{\sm} (001) surface}
\textit{(001) Surface terminations.}
%In order to verify the topological properties of a material, it is essential to study its surface.
{\sm} crystallizes into a simple cubic lattice, with lattice constant $\aaa=4.13$\AA. Crystals cleave preferentially along the (001) direction.
This surface, however, presents many challenges from both experimental and theoretical point of view, since it can come with many different terminations
(Sm\cite{hoffman_smb6,pnas_smb6_stm,smb6_stm_china}, B$_6$\cite{pnas_smb6_stm}, Sm reconstructed\cite{hoffman_smb6,pnas_smb6_stm}, single B\cite{smb6_stm_china}, six or eight B donut-like\cite{smb6_stm_china},
disordered\cite{hoffman_smb6,pnas_smb6_stm}).
All of these have different electronic properties, and are both difficult to control experimentally and to describe in a coherent fashion theoretically.

In particular, simple Sm and B$_6$ terminated surface are polar, which has lead to the suggestion that the observed surface conductance might be carried by non-topological surface states \cite{sawatzky_smb6}.
STS experiments \cite{pnas_smb6_stm,hoffman_smb6}, however, show that, while these nominally polar surface regions can be observed, they are somewhat rare and small, being electrostatically unstable.
The most common situation is instead the one of a reconstructed surface, in which, on average, one half of the terminating atoms are missing, restoring electrostatic neutrality.

\begin{figure}[tb]
\includegraphics[width=0.42\textwidth]{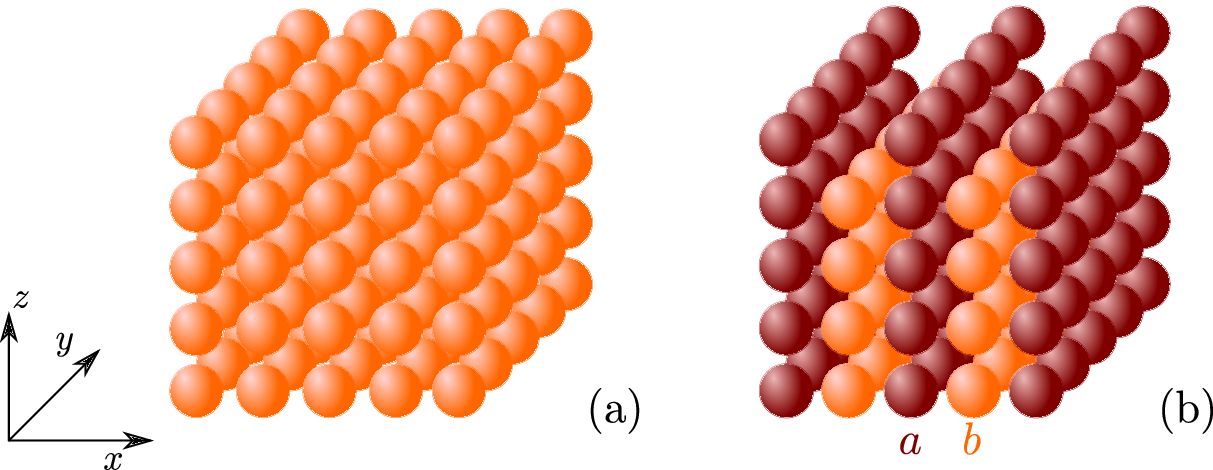}
\caption{Pictorial view of the (a) non-reconstructed (001) surface and (b) of the $2\times 1$ reconstructed one;
we only show Sm atoms, and in (b) we highlight with two different colors the two sublattices $a$ and $b$.
}
\label{fig_001}
\end{figure}

\textit{$2\times 1$ reconstructed (001) surface.}
Even though this reconstruction is most often disordered, large regions of ordered, Sm terminated, $2\times 1$ reconstructed (001) surfaces can be observed\cite{pnas_smb6_stm,hoffman_smb6}. In these areas every second Sm row on the top layer is missing, Fig.~\ref{fig_001}.
Due to the absence of macroscopic surface charges, this is the experimental scenario which most closely resembles the idealized (001) surface
which is usually described by tight-binding-based theoretical models\cite{lu_smb6_gutz,smb6_korea2,pub6,prbr_io_smb6,takimoto,tki_cubic,smb6_tci_io}.

\textit{Impact of reconstruction on surface states.}
The removal of one half of the atoms in the top layer cannot be considered a harmless process;
this is especially true if we consider surface states, which mostly live in the few top layers.
Due to relaxation effects and the removal of neighboring atoms, even layers below the reconstructed one are expected to be influenced by the process.
Here, the penetration depth $\lambda$ of surface states plays an important role:
if $\lambda/\aaa\gg 1$, these states live on many planes, and a strong perturbation on one or a few of them has a weak impact;
if instead $\lambda/\aaa\sim 1$, they live on just a few top layers, and a strong perturbation there can have a relevant effect.
Using a simple model
\cite{smb6_dzero_pert},
it has been found that the penetration depth is of the order of a few atomic layers,
suggesting the second possibility to be relevant.

\subsection{Modeling}

Even though it is generally agreed that the destruction of one or more of the top atomic layer(s) via disorder simply shifts topological surface states
deeper in the bulk\cite{fehske,smb6_iondamaged}, not much is known about a strong but periodic perturbation, as the $2\times 1$ reconstruction that we consider here.
It is therefore the aim of this paper to discuss the effect of this scenario on the electronic surface states. To this end, we shall employ an eight-orbital tight-binding model,
following Refs. \onlinecite{tki_cubic,prbr_io_smb6,smb6_tci_io} (see the Appendix for details).

\textit{Bulk mean-field approach.}
We recall that {\sm} is a strongly interacting material, in which local $f$ moments are coherently screened by conduction $d$ electrons to lead to a Kondo insulating phase. At low temperatures, this physics can be captured using a single-particle description with renormalized parameters \cite{hewson}:
in particular, the $f$ kinetic energy is strongly suppressed, by a factor that dynamical mean-field theory (DMFT) and the Gutzwiller approximation find to be $\sim 10$\cite{lu_smb6_gutz,pub6,smb6_korea2}.
Simpler theories such as mean-field slave bosons \cite{hewson} can qualitatively reproduce this reduction, but with a smaller renormalization factor $\sim 2$\cite{tki_cubic,prbr_io_smb6}.
Instead of describing interaction-induced renormalizations explicitly, here we will work directly with renormalized parameters \cite{renorm_foot} that are designed to match qualitatively the computed bandstructure and the measured ARPES dispersion.

\textit{Role of surfaces and Kondo breakdown.}
In presence of spatial inhomogeneities, the interaction-induced renormalizations will be position-dependent\cite{prb_io_tki}.
In particular, Kondo screening can be appreciably reduced on a surface, leading to an effective decrease of the Kondo temperature.
This can generate a so-called ``Kondo-breakdown'' \cite{kondo_breakdown},
in which surface moments are no more Kondo screened: they can be free, order magnetically, or be possibly still screened but with a Kondo temperature smaller than the bulk one.
Decoupled moments lead to an increase of the Dirac velocity, and to a displacement of the Dirac energy deep in the valence band, which resembles experimental results indicating ``light'' rather than ``heavy'' surface states.
In Ref. \onlinecite{kondo_breakdown} the Kondo temperature was estimated to drop by roughly a factor 3, as a consequence of surface Sm atoms having one first-nearest neighbor (NN) Sm atom less than in the bulk.
In presence of a $2\times 1$ reconstruction atoms on the top row will have three neighbors less:
in this case, the reduction of the hybridization, hence of the Kondo temperature, at the surface would be even larger.

However, an analysis \cite{prbr_io_smb6,smb6_tci_io} of tight-binding models built to fit ab-initio calculations indicates that the bulk gap is mostly generated by 2nd NN hybridization,
and in Ref. \onlinecite{smb6_broholm_exciton} a model with 3rd NN hybridization only seemed to reproduce well neutron-scattering data. This then leads to a more complex scenario, where the reduction of the hybridization of the surface atoms is less drastic.
We summarize the count of different NNs in Table \ref{table_nn}. While we will not model Kondo breakdown explicitly, we will return to this aspect in the next paragraph.

\textit{Additional surface scattering potential.}
To formally eliminate half of the atoms on layer $1$, we introduce a virtually infinite reconstruction potential $\Vrec$ acting every second row (in practice we take $\Vrec=100$\,eV).
With $\Vrec$ being active, we need to introduce an additional surface scattering potential $\Delta E=\Delta E_d=\Delta E_f$ of a few hundred meV, acting on the $d$ and $f$ orbitals of the remaining atoms in layer $1$, in order to qualitatively reproduce the surface-state dispersion as seen by ARPES (large Dirac velocity, Dirac energy in the valence band).
For $f$ electrons, which have a bandwidth of a few meV, this value turns out to be very close to the infinite scattering potential of the Kondo-breakdown scenario $\Delta E_f=\infty$,
in which surface $f$ electrons are no more part of the coherent Fermi liquid.
Even though quantitative details might vary between our approach and the Kondo breakdown one,
most of the general features we are going to describe are general, and independent of microscopic details.
We also remark that, while a large $\Delta E_f$ is needed to reproduce light surface states, $\Delta E_d$ is introduced to further adjust the position of the Dirac energy, hence the Fermi momentum $k_F$ (that we try to obtain as close as possible to experiments), and may not be needed for some choices of the parameters (see Section~\ref{sec:num_other}).

%\textit{Effective reconstruction scattering potential (ERP).}
%For example, both the scenarios we just described should entail a reduction of the (ERP) as compared to the simple estimation from the penetration length $\lambda$,
%since surface states will now have from the beginning a smaller weight on the first atomic layer,
%just restricted to $d$ orbitals in the limiting case of Kondo breakdown.

\begin{figure}[tb]
\includegraphics[width=0.45\textwidth]{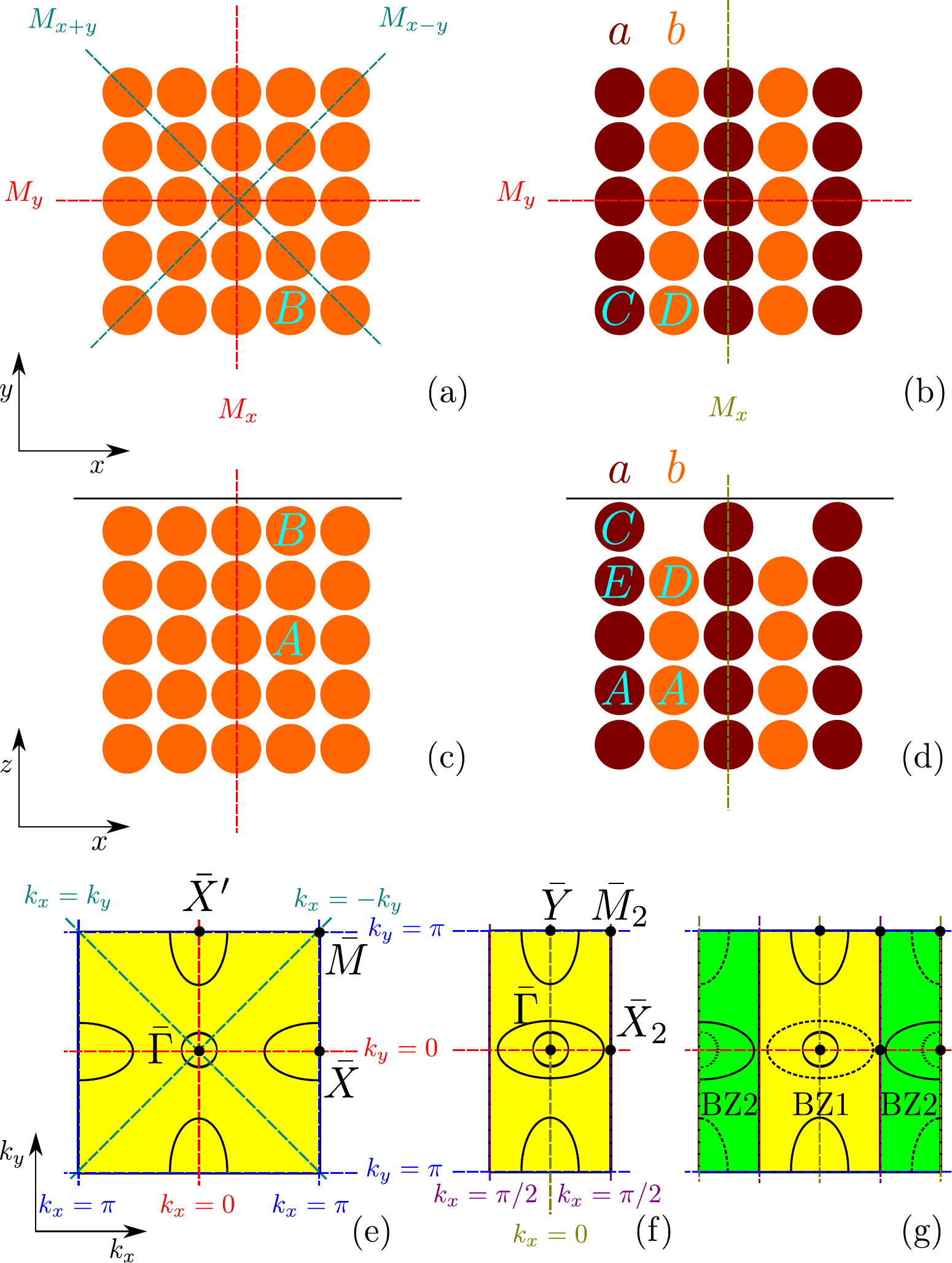}
\caption{Sketch of the Sm sites for a (001) surface (a)-(c) without reconstruction and (b)-(d) with a $2\times 1$ reconstruction.
In the first case, of $C_{4v}$ symmetry, four mirror planes are present (corresponding to operators $M_{x}$, $M_{y}$, $M_{x-y}$, $M_{x+y}$), while in the second case, of $C_{2v}$ symmetry, only $M_{x}$ and $M_{y}$ remain;
in this second case we depict the two sublattices with different colors.
Corresponding two-dimensional Brillouin zones (e) without reconstruction (``large'' BZ) and (f)-(g) with reconstruction;
in this second case the BZ is shrunk along the $x$ direction (``small'' BZ), and the $\bar X$ cone is folded at $\bar\Gamma$ (f);
however, for weak reconstruction, it is useful to use the original large BZ (g), with weak band replicas appearing as a consequence of backfolding (dotted ellipses).
In the non-reconstructed case high-symmetry points are labeled as $\bar\Gamma=(0,0)$, $\bar{X}=(\pi,0)$, $\bar{X}'=(0,\pi)$ and $\bar{M}=(\pi,\pi)$, and mirror planes are ${k_x=0}$, ${k_y=0}$, ${k_x=\pi}$, ${k_y=\pi}$, ${k_y=+k_x}$, ${k_y=-k_x}$,
which are pairwise equivalent by $C_{4v}$ symmetry.
In the reconstructed case, high-symmetry points are labeled as $\bar\Gamma=(0,0)$, $\bar{X_2}=(\pi/2,0)$, $\bar{Y}=(0,\pi)$ and $\bar{M}_2=(\pi/2,\pi)$,
and mirror planes are ${k_x=0}$, ${k_y=0}$, ${k_x=\pi/2}$, ${k_y=\pi}$.
%We also draw the 2D Fermi surface of \sm, consisting of three pockets, centered for the unreconstructed case at $\bar\Gamma$ and at the two $\bar X$ points;
%in the reconstructed case, the cone at $\bar X$ is folded at $\bar \Gamma$.
}
\label{fig_bz}
\end{figure}

\begin{table}
\centering
$$
\begin{array}{|l|c|c|c|}\hline
\mbox{Atom} &\mbox{1{st} NN} &\mbox{2{nd} NN} &\mbox{3{rd} NN}\\\hline
A\mbox{(bulk)}& 6 &12 &8\\
B\mbox{(non-rec. surface)}&5&8&4\\
C\mbox{(rec., top row)} &3 &4&4\\
D\mbox{(rec., below the missing row)}&5 &10&8\\
E\mbox{(rec., below the existing row)}&6 &10&4\\\hline
\end{array}$$
\caption{Number of Sm neighbors for Sm atoms (A) in the bulk, (B) on the (001) non-reconstructed surface, (C) in the first [(D,E): second] row on the $2\times 1$ reconstructed (001) surface; different positions $A$-$E$ are illustrated in Fig.~\ref{fig_bz}(a-d).
 }\label{table_nn}
\end{table}

\subsection{Qualitative effects of reconstruction}

In the following we choose a coordinate system as in Fig.~\ref{fig_bz}, with $\hat{z}$ perpendicular to the surface and $\hat{y}$ being parallel to the surface rows.

\textit{Surface Brillouin zone.}
A $2\times 1$ reconstruction doubles the size of the primitive surface unit cell in $\hat{x}$ direction and introduces two sublattices, $a$ and $b$ with and without top row, respectively, Fig.~\ref{fig_bz}(a-d). The two-dimensional (2D) surface BZ, originally a square for the (001) surface, shrinks by one half in $\hat{x}$ direction according to the reconstruction wavevector $\bQ=(\pi,0)$, Fig.~\ref{fig_bz}(e-f). Here we will use the term ``2nd BZ'' for the difference between the original (``large'') and the folded (``small'') BZ.

\textit{Spectral weights.}
A (weak) reconstruction causes a backfolding of bands which leads to replicas (or shadow bands) shifted by $\bQ$ w.r.t. the original bands and with reduced spectral weight. Experimentally, a weak replica of the $\bar X$ cone has been observed at $\bar\Gamma$\cite{mesot_smb6,smb6_arpes_feng} in ARPES,
even though the origin of this signal is not totally clear (for example, having polycrystalline samples, one would expect to observe the replica of the $\bar X'$ cone at $\bar\Gamma$, too).

Within our model calculations, we can quantitatively define spectral weights in the following way. The wavefunction $\psi_n(\bk,z,\alpha,m)$ of a single-particle state with energy $E_n$ for a system in slab geometry depends on the momentum $\bk\equiv (k_x,k_y)$ in the small BZ ($-\pi/2\le k_x \le \pi/2$, $-\pi\le k_y \le \pi$), on the layer index $z$, on the orbital index $\alpha$, and on the sublattice index $m=0(a),1(b)$.
Its spectral weight in the large BZ is given by:
\be\label{w_largeBZ}
w(n,\bK,z,\alpha)=\left|\sum_{m=0,1}e^{\ii K_x m}\psi_n(\bk,z,\alpha,m)\right|^2,
\ee
with $\bK\equiv (K_x,K_y)$ from the large BZ ($-\pi\le K_x,K_y \le \pi$), and $\bk$ its backfolded partner, $\bk=\bK\mod\bQ$.
We can define a global (i.e. orbital-, layer-, and momentum-integrated) energy-resolved weight separately for the 1st or 2nd BZ:
\bea
w^{BZ1}(E)=\sum_{n,\bK^{1st},z,\alpha}w(n,\bK^{1st},z,\alpha)L(E-E_n), \label{eqbz1}\\
w^{BZ2}(E)=\sum_{n,\bK^{2nd},z,\alpha}w(n,\bK^{2nd},z,\alpha)L(E-E_n), \label{eqbz2}
\eea
where $L$ is a Lorentzian kernel, and $\bK^{1st}$ means $|K_x|\le \pi/2$, while $\bK^{2nd}$ means $|K_x|> \pi/2$.
We can also define a layer-resolved weight $w_z$,
\be
%w_z(E)&=&\sum_{n,\bk,\alpha,m}|\psi_n(\bk,z,\alpha,m)|^2 L(E-E_n)=\\
w_z(E)=\sum_{n,\bK,\alpha}w(n,\bK,z,\alpha) L(E-E_n),
\ee
and decompose it into contribution from the 1st and 2nd BZ by limiting the sum over $\bK$ accordingly, to yield
$w_z^{BZ1}(E)$ and $w_z^{BZ2}(E)$, with $w_z^{BZ1}(E)+w_z^{BZ2}(E)=w_z(E)$, and
$\sum_z w_z(E)=\rho(E)=w^{BZ1}(E)+w^{BZ2}(E)$, where $\rho(E)=\sum_n L(E-E_n)$ is the density of states.

\textit{Effective reconstruction potential.}
To quantify the effect of reconstruction on surface states, we introduce the qualitative notion of ``effective reconstruction potential'' (ERP) strength which is based on spectral weights. ERP is a dimensionless quantity between 0, implying no effect, and 1, implying maximal effect.

We start from an idealized situation, in which the reconstruction potential only affects wavefunctions in the first layer $z=1$. On this layer, the weight in the 1st and 2nd BZ are equal, $w(n,\bK^{1st},z=1,\alpha)=w(n,\bK^{2nd},z=1,\alpha)=w(n,\bK,z=1,\alpha)/2$, as only the $m=0$ term in Eq. \eqref{w_largeBZ} survives. In all other layers, the weight (of a single state) will be totally either in 1st or in the 2nd Brillouin zone, according to where it was without reconstruction.
Hence, the ratio of the layer-integrated weights in the 1st and 2nd BZ (i.e. the relative weight of replicas) can be used to infer the weight of surface states on the top layer. Specifically, for a band located in the 2nd BZ without reconstruction (like the $\bar X$ cone), with $w(n,\bK^{1st},z>1,\alpha)=0$, $w(n,\bK^{2nd},z>1,\alpha)=w(n,\bK,z>1,\alpha) \ne 0$, the normalized weight on the first layer can be written as:
\be\label{eq_erp}
w_{1}'\equiv\frac{w_1}{\rho}=\frac{2w^{BZ1}}{w^{BZ1}+w^{BZ2}}, \hspace{5pt} 0\le w_1'\le 1.
\ee
%If the first scenario is right, and we can assume a simple wavefunction of the form $\psi(z)=\exp(-z/\lambda)/\lambda$,
%this can be used to extract to penetration depth $\lambda$ of surface states:
%$\lambda=1/(2w_1)$,
%where the factor 2 comes from the fact that on the first layer the weight is reduced by the absence of half of the atoms;
%but this formula fails in the second scenario, where the strong surface scattering potential must modify $\psi(z)$ for small $z$,
%even though the asintotic exponential behavor is described by the same $\lambda$.
We note that the above assumption is not strictly valid: We found that the weight of the $\bar X$ cone
can be non-negligible in the 1st BZ even for deeper layers $z>1$, even though no explicit reconstruction potential is present there, see Section~\ref{sec_numerics}. Nevertheless, we shall use $w_{1}'$ from Eq. \eqref{eq_erp}, evaluated for a surface state at the Fermi level, as a measure of the ERP. We will see in Section \ref{sec_numerics} that this value can also be cone-dependent.

\textit{How strong is the reconstruction?}
We now briefly discuss whether existing experimental or theoretical data are sufficient to infer the effective strength of the reconstruction potential.

Experimentally, ARPES data indicate that band replicas are weak and hence the ERP, as measured by Eq. \eqref{eq_erp}, is small\cite{mesot_smb6,smb6_arpes_feng}. This is, however, difficult to reconcile with our model calculations which, under essentially all circumstances, show that the removal of every second Sm atoms in the surface layer is a strong perturbation to the surface states, with large ERP (see Section \ref{sec_numerics} for details).
Possible reasons for this discrepancy are:
(i) only a fraction of the surface probed by ARPES features an ordered $2\times 1$ surface reconstruction, thus reducing the ERP in the collected ARPES signal, or
(ii) surface states in the real material tend to be expelled from the first layer(s) by some other mechanism, such as Kondo breakdown, or
(iii) the penetration length of surface states is too small in the model calculations.
Issue (i) may be investigated with small-spot ARPES experiments, and we will not discuss this further.
For (ii), we found that the introduction of the surface scattering potential, $\Delta E_d$ and $\Delta E_f$, to model this expulsion often enhances rather than decreases the ERP, but this is a model-dependent statement. We note that ARPES may not probe the full weight of surface states if their penetration length is larger than that of the ARPES probe, but this is unlikely to explain the above discrepancy.
The issue (iii) of correctly modelling the surface-state penetration is delicate: The problem of the excessive reduction of the bulk gap using renormalized parameters (with respect to plain DFT ones, noted for example in Ref. \onlinecite{smb6_denlinger_t})
has a direct impact here: if one wants to match the experimental ARPES gap $\sim 20$ meV using renormalized parameters, one has to increase the value of the hybridization, effectively decreasing the penetration length of surface states, and, as a consequence, enhancing the strength of the ERP.
The problem is further complicated by $\Gamma_7$ states\cite{pub6,prbr_io_smb6,smb6_tci_io,smb6_korea2},
whose inclusion would reduce the value of the bulk gap (in our model, we will ignore these states, but a more detailed analysis would need to include them).

We conclude that we are not in the position to fully resolve this discrepancy, and further work -- both theoretical and experimental -- is needed to do so.
Instead, we will take a pragmatic point of view and analyze the fate of the surface states provided that the ERP is sufficiently strong to generate observable effects.

%%%%%%%%%%%%%%%%%%%%%%%%%%%%%%%%%%%%%%%%%%%%%%%%%%%%%%%%%%%%%%%%%%%%%%%
%%%%%%%%%%%%%%%%%%%%%%%%%%%%%%%%%%%%%%%%%%%%%%%%%%%%%%%%%%%%%%%%%%%%%%%
%%%%%%%%%%%%%%%%%%%%%%%%%%%%%%%%%%%%%%%%%%%%%%%%%%%%%%%%%%%%%%%%%%%%%%%

\section{Surface states with reconstruction}\label{sec_rec}

In this Section we qualitatively analyze the effect of surface reconstruction on the in-gap surface states, based mainly on symmetry arguments. Numerical examples, taking into account details of microscopic modelling and testing these predictions, will be provided in Section~\ref{sec_numerics}.

\subsection{Symmetries and dispersion}

\textit{Energies close to the Dirac point.}
If the Dirac energies were close to the Fermi level, the Fermi wavevector $k_F$ of the cones were small, and essentially the only effect of reconstruction would be the folding of bands into the small BZ, as shown in Fig.~\ref{fig_bz}(f)-(g). (The same applies at all energies for weak ERP.)

\textit{Overlap of $\bar X$ cone with its replica.}
Away from the Dirac energies, Dirac cones and their replicas will start to overlap with each other. ARPES data on {\sm} show that the only relevant overlap is the one of the $\bar X$ cone with its replica: $2k_F$ of the $\bar{X}$ cone along the $\bar{X}-\bar\Gamma$ direction is estimated to be\cite{mesot_smb6} around $0.80$\,\AA$^{-1}$, while the original BZ has a halfwidth $\pi/\aaa=0.76$\,\AA$^{-1}$. Thus, we expect that, due to the reconstruction, the $\bar{X}$ cone should overlap and hybridize with its replica for energies close to the Fermi energy; we will ignore similar effects which might affect the $\bar X'$ (now $\bar Y$) and $\bar\Gamma$ cones at higher energies.

\begin{figure}[tb]
\includegraphics[width=0.49\textwidth]{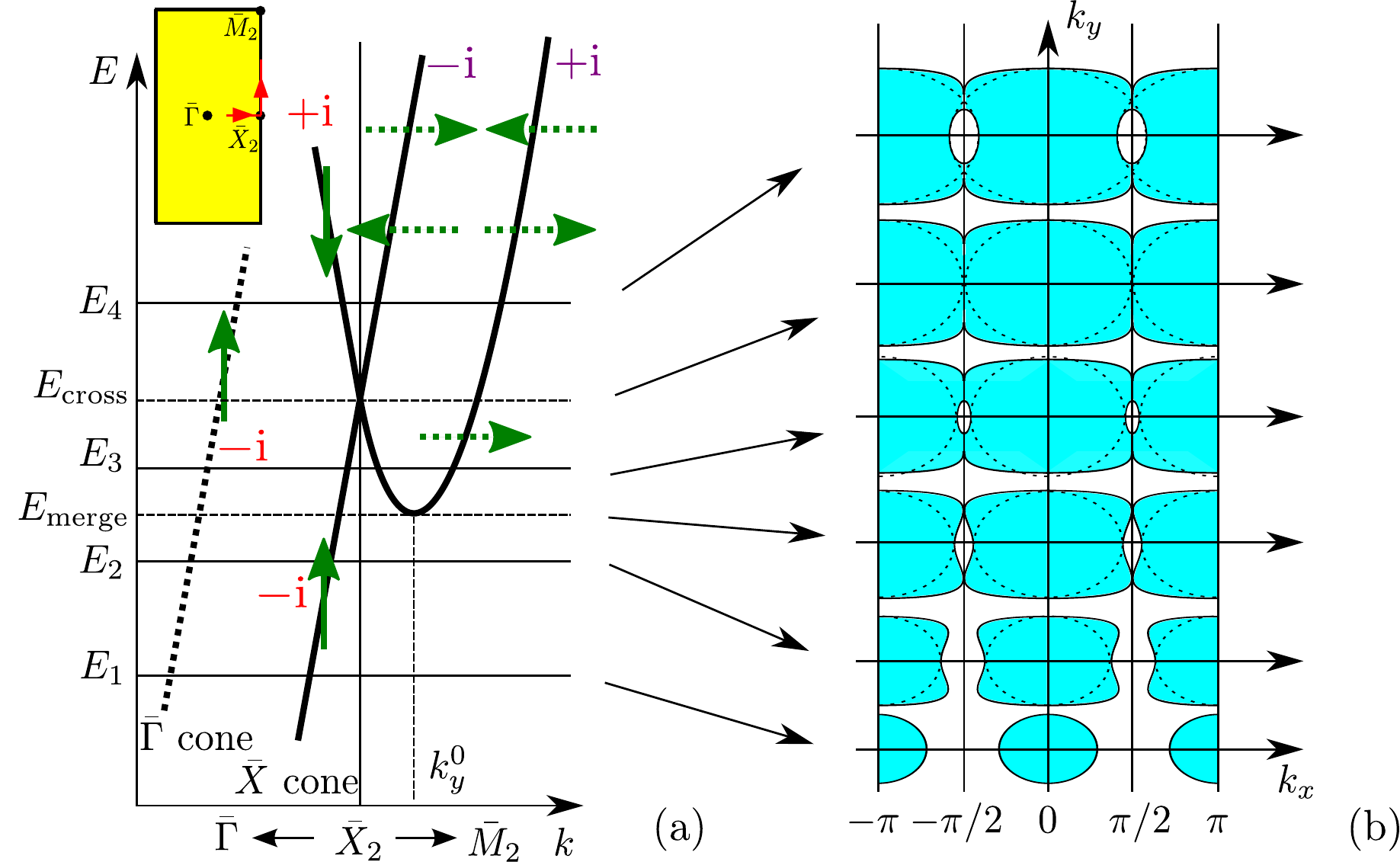}
\caption{Lifshitz transition for the $\bar{X}$ cone. %in the $(C^+_{k_z=0}, C^+_{k_z=\pi},C^+_{k_x=k_y})=(+2,+1,-1)$ phase.
(a) Dispersion along the $\bar\Gamma - \bar{X}_2 - \bar{M}_2$ lines (the $\bar\Gamma$ cone is a spectator).
At $E=\Emerge$ the $\bar{X}$ cone merges at $(k_x,k_y)=(\pi/2, \pm k_y^0)$ with its own replica, centered at $\bar\Gamma$, giving rise to the Lifshitz transition,
while at $E=\Ecross$ the two cones cross along the $k_y=0$ line: along this direction they cannot hybridize, having different mirror-symmetry $M_y$ eigenvalues.
We also report mirror eigenvalues $\pm\ii$, and draw the SEV with green arrows (for solid arrows the relation to mirror eigenvalues is univocal, for dotted ones is model dependent).
(b) Schematic evolution of the $\bar X$ cone and of its replica across the transition; with a dotted ellipse we show how the cones would evolve without reconstruction potential $\Vrec$.
}
\label{fig_lifshitz}
\end{figure}

\textit{Hybridization, role of mirror symmetry, and Lifshitz transition.}
We now discuss general aspects of the coupling between the $\bar X$ cone and its replica which are largely independent of microscopic details; see Fig.~\ref{fig_lifshitz}. A key ingredient is the reflection symmetry with respect to the $xz$ plane, described by operator $M_y$: States with $k_y=0$ are invariant under $M_y$ (Ref.~\onlinecite{smb6_tci}). Along this direction the two cones have the smallest distance, and the overlapping states have opposite mirror eigenvalues $\pm \ii$ (their exact value depends on the sign of the mirror Chern number $\C^+_{k_z=0}$ but is irrelevant for the argument). Hence, the cones cannot hybridize for $k_y=0$ (as long as mirror symmetry is preserved), instead they cross at $E=\Ecross$.

In contrast, hybridization is allowed for $k_y\neq 0$. We observe that for energies $\Ek$ corresponding to momenta $k\lesssim \pi/2$, the $\bar X$ cone deforms ($E=E_1$ in Fig.~\ref{fig_lifshitz}) along the $k_y=0$ direction, until a
Lifshitz transition at $E=\Emerge<\Ecross$ is achieved: the $\bar X$ cone and its replica merge at $(\pi/2,k_y^0)$, where in general $k_y^0\ne 0$.
As a consequence, a pocket centered at $\bar X_2=(\pi/2,0)$ is created;
we note that the minimum along $\bar X_2$ - $\bar M_2$ is actually a saddle point.
When the two cones cross at $E=\Ecross$ along the $k_y=0$ direction, the $\bar X_2$ pocket area vanishes. At higher energies $E>\Ecross$, the $\bar X_2$ pocket grows again, such that a new Dirac cone is created at the $\bar X_2$ pocket from the crossing of the two original Dirac cones.
We note that for vanishing scattering potential, the two cones always cross without hybridizing: this yields $\Emerge=\Ecross$, and the two branches along the $\bar X_2 - \bar M_2$  direction ($k_x=\pi/2$) are degenerate. Finally, for the case with reconstruction and {\em broken} mirror symmetry, a gap is opened at $\bar X_2$.

\textit{Analogy to TCIs.}
The hybridization behavior in Fig.~\ref{fig_lifshitz} is similar to what happens on the (001) surface of face-centered-cubic TCIs such as Pb$_{1-x}$Sn$_x$Se\cite{pbsnse_tci}, Pb$_{1-x}$Sn$_x$Te\cite{pbsnte_tci}, and SnTe\cite{snte_tci}.
For these materials, for which the standard $\mathbb{Z}_2$ indices are zero and which are not TIs in the usual sense, mirror Chern numbers are non-zero, and predict two pairs of Dirac cones along the $k_x=0$ and $k_y=0$ directions of the 2D BZ. These cones are at $(\pi\pm\delta_k,0)$ and $(0,\pi\pm\delta_k)$, where the BZ is defined by $-\pi<k_x,k_y<\pi$.
Each pair of cones merge around $(\pi,0)$ or $(0,\pi)$ through the same mechanism that we have described here, and which has been also predicted\cite{smb6_tci} for the (110) surface of {\sm}.
The similarity lies in the non-trivial topology of the band structure and the presence of mirror planes, which create states with different mirror eigenvalues which cannot hybridize along specific  directions.

However, in the TCI case, no surface reconstruction is needed to generate the pair of cones. Also, the role of symmetries is reversed: in standard TCIs mirror symmetry protects cones at $(0,\pi\pm\delta_k)$, while parity invariants predict two Dirac cones at $(0,\pi)$.
In our case we identify $(0,\pi)$ with $\bar X_2$ and $(0,\pi\pm\delta_k)$ with $\bar\Gamma$ and $\bar X$, but now mirror symmetry protects the cones at $\bar X_2$, while parity invariants predict cones at $\bar\Gamma$ and $\bar X$.

ARPES experiments on TCIs have verified \cite{pbsnse_tci,pbsnte_tci,snte_tci_arpes} the surface-state behavior as depicted in  Fig.~\ref{fig_lifshitz}. While a similar verification might be possible in {\sm}, we note a few caveats: Visible hybridization effects require the ERP to be strong; as noted in the previous Section its magnitude is unknown. Moreover, the features might be beyond the present ARPES energy and momentum resolution, and finally the cone crossing might occur above the Fermi energy, rendering ARPES ineffective.

\subsection{Direction of spin}

\begin{figure}[tb]
\includegraphics[width=0.48\textwidth]{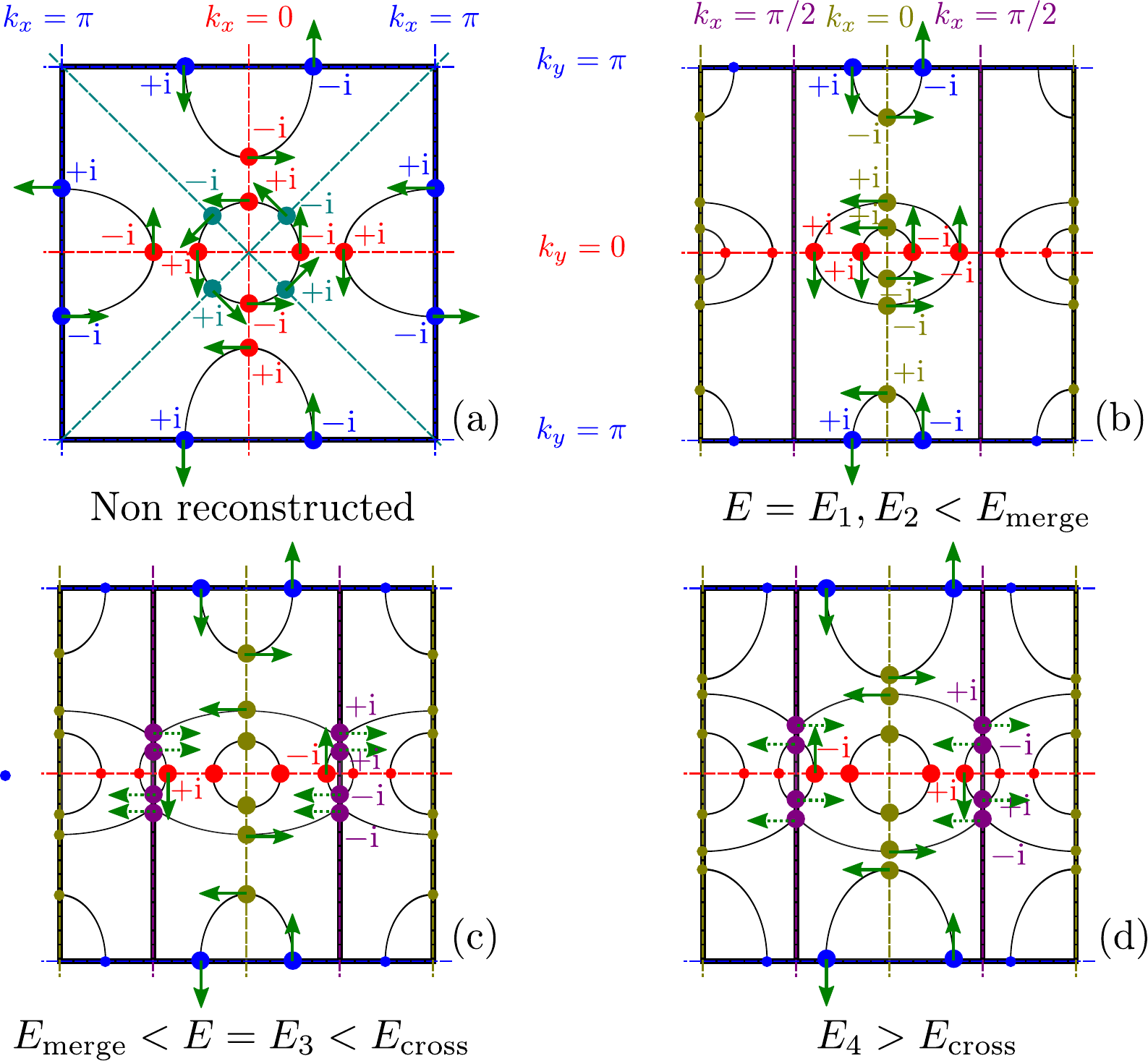}
\caption{Mirror-symmetry eigenvalues $\pm\ii$ and SEV for (a) the non-reconstructed and (b-d) the reconstructed BZ at different energies, see Fig.~\ref{fig_lifshitz}.
Mirror eigenvalues and SEV not reported in panels (c,d) are the same as in panel (b).
}
\label{fig_chern}
\end{figure}

So far, the analysis of crossing Dirac cones was based on the presence of a mirror plane, but was otherwise independent of microscopic details. We now take a step further and compute mirror-symmetry eigenvalues and the expectation value of the electronic spin operator.
As shown in Refs.~\onlinecite{kane_bisb, sigrist_tki_prb,smb6_tci_io,smb6_tci_legner}, mirror-symmetry eigenvalues are determined by the sign of mirror Chern numbers (see Appendix). While four TCI phases with $\C^+_{k_z=0}=\pm 2$, $\C^+_{k_z=\pi}=+1$, $\C^+_{k_x=k_y}=\pm 1$ are possible in principle, we restrict the analysis to the $\C^+_{k_z=0}=+2$, $\C^+_{k_z=\pi}=+1$, $\C^+_{k_x=k_y}=-1$ phase, which -- as we have argued in Ref.~\onlinecite{smb6_tci_io} -- is the one of experimental relevance for {\sm}.
In Fig.~\ref{fig_chern}(a) we report the mirror-symmetry eigenvalues for the non-reconstructed surface.\cite{smb6_tci_io}

%We now want to fix mirror eigenvalues for all possible states also in the reconstructed case.
\textit{Mirror-symmetry eigenvalues with reconstruction.}
In the reconstructed case, there are four distinct mirror planes in the 2D BZ, namely $k_x=0, \pi/2$ and $k_y=0,\pi$; the former commute with $M_x$, the latter with $M_y$ (see Fig.~\ref{fig_bz}(b)-(d)-(f)). We can therefore assign mirror eigenvalues to states crossing these planes.
For planes $k_x=0$, $k_y=0$, $k_y=\pi$ everything is equivalent to the non-reconstructed case, and we can transfer the eigenvalues directly from Fig.~\ref{fig_chern}(a) into Fig.~\ref{fig_chern}(b).
For energies below $\Emerge$, no states cross $k_x=\pi/2$, and no further work is needed.
Above $\Emerge$, however, four states are crossing the $k_x=\pi/2$ plane, two for $k_y>0$, and two for $k_y<0$. Concentrating on $k_y>0$ we find that below $\Ecross$ both states have eigenvalue $-\ii$, see Fig.~\ref{fig_chern}(c), simply because they belong to the same band, see Fig.~\ref{fig_lifshitz}(a). In contrast, above $\Ecross$, Fig.~\ref{fig_chern}(d), one of the $k_y>0$ states has eigenvalue $+\ii$, and the other $-\ii$, see Fig.~\ref{fig_lifshitz}(a), since they now belong to different bands;
the same happens for $k_y<0$, see Fig.~\ref{fig_chern}(c,d).

\textit{Spin expectation value (SEV).}
Since mirror-symmetry eigenvalues are not directly measurable, we now concentrate on the SEV, which depend on these eigenvalues and is accessible in SP-ARPES experiments.
For atomic $d$ and $\Gamma_8$ states, that we consider here, the relation between mirror eigenvalues and SEV is straightforward, and it is the same one as for free spins\cite{smb6_tci_io}:
if the eigenvalue is $-\ii$, the SEV points along $+x$ for $M_x$ and $+y$ for $M_y$, and viceversa for eigenvalues $+\ii$.

This holds also for lattice states with a definite momentum, when symmetry operations act like in the non-reconstructed case. This applies to states $k_x=0$, $k_y=0$, or $k_y=\pi$ -- hence, on these planes, the SEV is identical to the non-reconstructed case -- but not for states at $k_x=\pi/2$ because this is not a BZ mirror plane in the absence of reconstruction.

Let us analyze the latter case of $k_x=\pi/2$ in more detail. For the two sublattices $a$ and $b$ we define mirror eigenstates according to $M_x|a^+\rangle=+\ii|a^+\rangle$, $M_x|b^+\rangle=+\ii|b^+\rangle$, and we build Bloch states $|\bk_a^+\rangle$ and $|\bk_b^+\rangle$ with $\bk=(\pi/2,k_y)$.
When we act with the mirror operator $M_x$ on $|\bk_a^+\rangle$ we get $M_x|\bk_a^+\rangle=+\ii|\bk_a^+\rangle$, but for states on sublattice $b$ we pick up a minus sign:
$M_x|\bk_b^+\rangle=-\ii|\bk_b^+\rangle$ [the opposite would happen if we took a mirror operator $M_x$ centered on $b$ rather than on $a$ in Fig.\ref{fig_bz}(b-d)].
This can be realized by observing that $M_x$, when centered on sublattice $a$, sends states $a$ in the $n$ primitive cell into states $a$ in cell $(-n)$,
while $b$ states at $n$ are sent into $b$ states at $(-n-1)$: so, the effect of $M_x$ on $|\bk_b\rangle$ states carries an additional $e^{-2\ii k_x}$ factor, which is $-1$ for $k_x=\pi/2$.
As a consequence, states $|\bk\rangle$ with $M_x|\bk\rangle=+\ii|\bk\rangle$  and $\bk=(\pi/2,k_y)$ are composed by states $|a^+\rangle$ on the sublattice $a$, and by states $|b^-\rangle$ on sublattice $b$,
so there is no general rule for their SEV given the mirror eigenvalues.
We can however state that, if we restrict the analysis to the top layer, where no $b$ states are present, eigenvalues $-\ii$ corresponds to a SEV pointing along $+\hat x$, and viceversa for states with $+\ii$.
When including all layers, however, we have to resort to numerical diagonalization.
In our model we find that, close to the Lifshitz transition, the total contribution from other layers is opposite to the one from layer $1$, so the SEV for $-\ii$ states points along $-\hat x$;
at higher energies, however, we observe a transition to a situation where the SEV for $-\ii$ states points along $+\hat x$ (see Fig.~\ref{fig_lifshitz}(a)), in agreement with the results from layer $1$.

%$\eta_x^{d1}=0$,
%$\eta_z^{d1}=0.8$,
%$\eta_z^{d2}=-0.3$
%$\eta_x^{f1}=1$
%$\eta_z^{f1}=-4$
%$\eta_z^{f2}=2$
%$\eta_x^{v1}=0.4$
%$v_1=-2$
%$v_2=0.6$

%%%%%%%%%%%%%%%%%%%%%%%%%%%%%%%%%%%%%%%%%%%%%%%%%%%%%%%%%%%%%%%%%%%%%%%
%%%%%%%%%%%%%%%%%%%%%%%%%%%%%%%%%%%%%%%%%%%%%%%%%%%%%%%%%%%%%%%%%%%%%%%
%%%%%%%%%%%%%%%%%%%%%%%%%%%%%%%%%%%%%%%%%%%%%%%%%%%%%%%%%%%%%%%%%%%%%%%

\section{Numerical Examples and QPI}\label{sec_numerics}

In this section we provide numerical examples, using explicit tight-binding calculations, of the scenario depicted in the previous Section, together with simulated QPI patterns.

\subsection{Introduction to QPI}
\textit{Generalities.}
QPI can be used as a probe of the topological character of surface states\cite{xue,yazdani,guo_franz_sti}, owing to their spin texture: non-magnetic impurities cannot induce transitions between states which are Kramers partners.
Also, when mirror symmetries are present, impurities which do not break these symmetries cannot induce transitions between states with opposite mirror eigenvalues\cite{tci_qpi_theory}.
In the following, we will exclusively consider impurities which do not break time-reversal and mirror symmetries; more complicated impurities may induce additional QPI peaks\cite{guo_franz_sti,tci_qpi_theory} which we will not discuss here.

In general, peaks in the QPI signal are expected for pairs of stationary states\cite{liu_qpi_spa} (states which have the same tangent in the BZ to the isoenergy contour), for which no selection rule forbids an elastic scattering process.
%Also, if the scattering wavevector are close to zero or correspond to a rigid shift of the isoenergy contour, phase space arguments greatly reduce its QPI weight (?).
For TI surface states, no peaks are expected from intracone scattering as long as the dispersion is linear, since pairs of stationary points are time-reversal conjugate in the Dirac Hamiltonian\cite{guo_franz_sti}. Peaks can arise when the dispersion deviates from linear, as it is well known to happen in TIs such as Bi$_2$Ti$_3$ due to hexagonal warping\cite{xue}.

\textit{Intercone scattering and QPI on {\sm} non-reconstructed (001) surfaces.}
Intercone scattering is less restricted by selection rules and may hence give rise to strong QPI peaks\cite{prb_io_tki}. However, for the non-reconstructed (001) {\sm} surface, we showed in Refs.~\onlinecite{prbr_io_smb6,smb6_tci_io} that the experimental spin pattern, corresponding to the
$\C^+_{k_z=0}=+2$, $\C^+_{k_z=\pi}=+1$, $\C^+_{k_x=k_y}=-1$ mirror Chern numbers and to a positive winding number $w\equiv\sgn(\C^+_{k_z=0}\C^+_{k_z=\pi})$ on the $\bar X$ cone, leads to a strong suppression of the intercone scattering, implying virtually no observable QPI peaks.
This most likely explains why STS experiments on {\sm} have not found noticeable QPI signals to date\cite{smb6_stm_china,hoffman_private_comm}.

%We anticipate that here mirror Chern numbers do not suffice to predict QPI patterns for each phase, since model details are important.
%We will try, however, to decribe what is general (for example, dependence on mirror eigenvalues and SEV), and what is model dependent (for example, orbital effects).

%We start with the unreconstructed case.
%By using this last property, we can immediately find prohibited transitions, which turn out to be independent from Chern numbers, and are illustrated in Fig.~\ref{fig_qpi_1}.
%They all involve intracone scattering, except for $\bar\Gamma$ - $\bar X$ intercone scattering along the $k_x=0$ or $k_y=0$ directions.

\textit{QPI in {\sm} with reconstruction.}
A key question is whether QPI peaks can appear as a consequence of surface reconstruction. As noted above, the question makes sense only for sufficiently strong ERP, otherwise the non-reconstructed picture, which shows no peaks, still holds to a good approximation.
We can expect new QPI scattering channels in the energy region around the Lifshitz transition:
first, because for energies below the transition an appreciable warping of the $\bar X$ cone is achieved, second, because above the transition the topology of the Fermi surface changes, with the appearance of the new pocket at $\bar X_2$. Concrete calculations are required to see whether these prerequisites yield a noticeable QPI signal -- those will be presented in what follows.

%As in Section \ref{sec_rec}, we will use the $\Gamma_8^{2z}$ model supplemented by the $v_1$ hybridization term, concentrating the analysis on the $\bar X$ cone and its replica.
%As shown above, this model has $w=-1$ and does not give any QPI peak in the unreconstructed case, since both intracone and intercone scattering are suppressed.

\subsection{Numerical results: Significant warping}

\begin{figure}[!tb]
\includegraphics[width=0.43\textwidth]{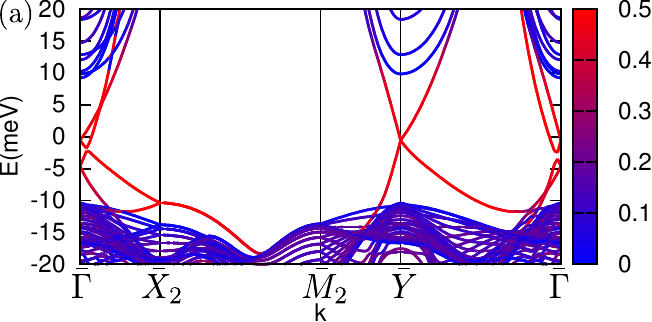}
\includegraphics[width=0.43\textwidth]{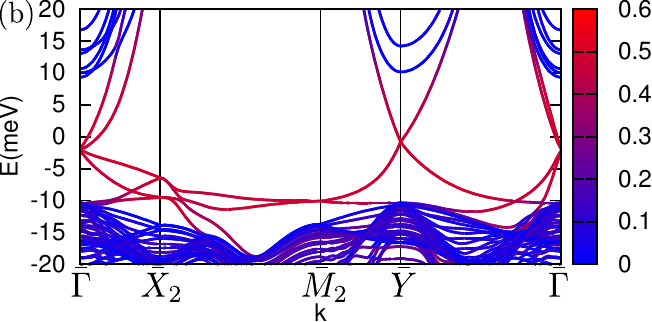}
\includegraphics[width=0.43\textwidth]{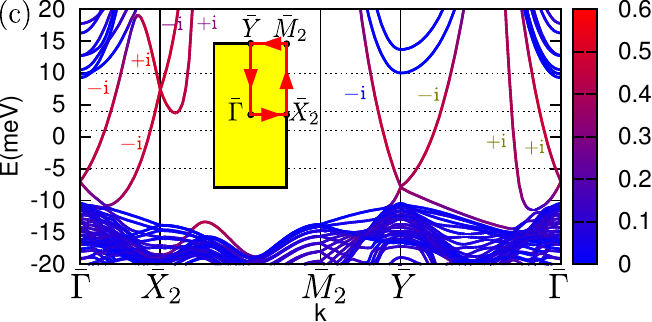}
\caption{Dispersion of a 15-layer slab along the $\bar\Gamma\bar X_2\bar M_2\bar Y\bar \Gamma$ path for
(a) $\Vrec=0$, $\Delta E=0$; (b) $\Vrec=100$\,eV, $\Delta E=0$; (c) $\Vrec=100$\,eV, $\Delta E=-0.5$\,eV.
The Fermi energy has been placed in the middle of the bulk gap.
Line colors encode the added weight on layers $1$ and $2$; for case (c) we also label surface states with mirror-symmetry eigenvalues $\pm\ii$.
We can observe two cones at $\bar \Gamma$ and one at $\bar Y$.
In (c), $\Emerge \approx 4$\,meV and $\Ecross\approx 7$\,meV, the latter corresponding to the Dirac-cone crossing at  $\bar X_2$.
Horizontal dashed lines correspond to the energies of the cuts in Fig.~\ref{fig_qpi}.
}\label{fig_bands}
\end{figure}

\begin{figure}[tb]
\includegraphics[width=0.44\textwidth]{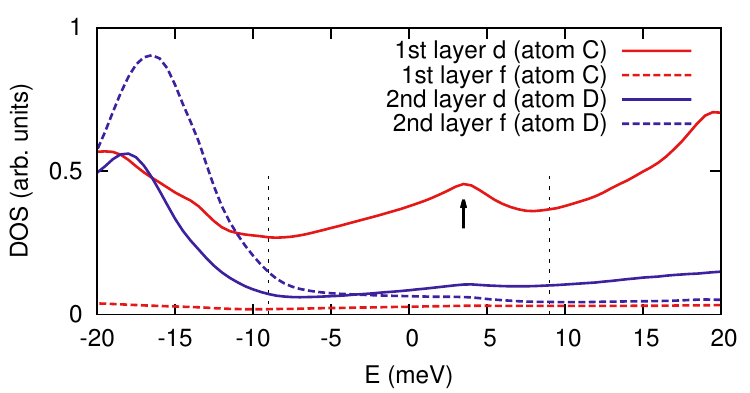}
\caption{Local DOS for $d$ and $f$ orbitals on atoms C and D from Fig.~\ref{fig_bz}, for parameters as in Fig.~\ref{fig_bands}(c).
The peak at $\Emerge\approx4$\,meV (arrow) originates from the Lifshitz transition;
the peak around $-17$\,meV is the bulk $f$ band. Dotted vertical lines around $\pm 9$\,meV denote the bulk band gap. The $f$ DOS on atom D is rescaled down by a factor 200.
}\label{fig_dos}
\end{figure}

\begin{figure}[!b]
\includegraphics[width=0.49\textwidth]{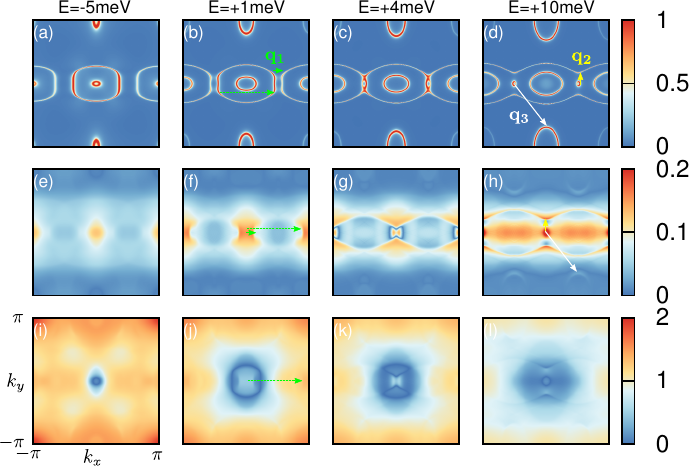}
\caption{
(a-d) ARPES signal from layers $1$ and $2$, Eq.~\eqref{arpes}, (e-h) QPI signal from layer $1$, and (i-l) QPI signal from layers $1$ and $2$ (divided by 10), all for parameters as in Fig.~\ref{fig_bands}(c).
Arrows illustrate scattering wavevectors.
Different columns correspond to different energies:
(a,e,i): far below the Lifshitz transition, yielding no coherent QPI signal;
(b,f,j): slightly below the transition, where a QPI signal at $\bq_1$ arises from warping (the solid and the dotted green arrows are equivalent up to an inverse lattice vector);
(c,g,k): at the transition;
(d,h,l): above the transition, with new scattering channels $\bq_{2,3}$ involving the new pocket at $\bar X_2$.
The QPI signal is normalized to the LDOS at the Fermi energy in the first layer.
}\label{fig_qpi}
\end{figure}

We employ the model described in the Appendix, taken from Refs. \onlinecite{prbr_io_smb6,smb6_tci_io}, which is designed to match as closely as possible the calculated bulk bandstructure, and to reproduce the experimentally observed surface-state dispersion and spin pattern. Without surface reconstruction
it yields no sharp QPI peaks, and consequently the appearance any QPI peaks can be traced to reconstruction effects.
For a specific set of parameters which produces significant warping near the cone crossing, we report the band dispersion for a slab geometry in Fig.~\ref{fig_bands}, the corresponding density of states (DOS) in Fig.~\ref{fig_dos}, and constant-energy cuts through the ARPES and QPI signals in Fig.~\ref{fig_qpi}. Here, we define the ARPES signal as the sum over the weights from all orbitals \cite{renorm_foot}, and add the signal from layers $1$ and $2$,
\begin{equation}
\label{arpes}
A(E,\bK) = \sum_{n,\alpha} \sum_{z=1,2} w(n,\bK,z,\alpha) L(E-E_n),
\end{equation}
expressed using the weight defined in Eq.~\eqref{w_largeBZ}.

\textit{Dispersion and DOS.}
In the dispersion, Fig.~\ref{fig_bands}, we can see two cones at $\bar \Gamma$, and one cone at $\bar Y$. For completeness, we present in Fig.~\ref{fig_bands}(a) the case without reconstruction, $\Vrec=0$, and in Fig.~\ref{fig_bands}(b) the case with reconstruction but no surface scattering potential, $\Delta E=0$. In what follows we mainly concentrate on Fig.~\ref{fig_bands}(c), which shows the case with both reconstruction and surface scattering potential.

The weight distribution in Figs.~\ref{fig_qpi}(a-d) shows that the $\bar Y$ cone is weakly affected by the reconstruction, and lives mostly in the 1st BZ: for this cone the ERP is weak.
In contrast, the two cones present at $\bar\Gamma$ are strongly hybridized with each other, and have appreciable weight both in the 1st and in the 2nd BZ even when considering the signal from deeper layers: for these cones the ERP is strong.
Their origin is from $\bar\Gamma$ and $\bar X$ in the non-reconstructed BZ; in analogy to the experimental situation we identify as ``$\bar\Gamma$'' the cone with Dirac energy close to the top of the valence band and smaller Fermi momentum,
and as ``$\bar X$'' the cone with lower Dirac energy and larger Fermi momentum (see Fig.~\ref{fig_lifshitz}).

We can observe that the $\bar X$ cone undergoes the Lifshitz transition that we described in the previous section, while the $\bar \Gamma$ cone remains inert. The saddle point at $\Emerge$ gives rise to a peak in the DOS (strictly speaking, a logarithmic divergence), as shown in Fig.~\ref{fig_dos}.
Parenthetically, we note that no cone crossing is observed inside the gap for significantly smaller values of the surface scattering potential $\Delta E$ (see for example Fig. \ref{fig_bands}(b)).

\textit{QPI.}
To illustrate QPI effects, we consider a single impurity (corresponding to the dilute limit) and employ the standard T matrix technique to compute the Fourier-transformed local DOS in the presence of elastic scattering; for details see the Appendix.
We report two sets of results: In Fig.~\ref{fig_qpi}(e-h) we show the QPI signal from layer $1$, which has equal weight in the 1st and in the 2nd BZ due to the missing $b$ atoms. In contrast, in Fig.~\ref{fig_qpi}(i-l) the signal is constructed from the sum of the local DOS in layers $1$ and $2$, the latter scaled down by a factor 10, to account for its larger distance from the tip. A more refined analysis might include interference contributions from $d$ and $f$ channels \cite{coleman09,prbr_io_smb6} as well as from layers $1$ and $2$, but would lead to the same qualitative results.

It can be observed that far from the Lifshitz transition no QPI signals, apart from an incoherent peak at $\bar\Gamma$, are present (Fig.~\ref{fig_qpi}(a));
this is the same result as without reconstruction.
At higher energies, slightly below the Lifshitz transition  a weak intracone scattering signal is generated as a consequence of cone warping (wavevector $\bq_1$ in Fig.~\ref{fig_qpi}(b)),
while above the transition two scattering channels appear to be active, both involving the newly formed $\bar X_2$ cone.
In the first one the scattering is towards the newly formed band coming from the merging of the $\bar X$ cone with its replica,
denoted $\bq_2$ in Fig.~\ref{fig_qpi}(d), and which involves states with the same mirror eigenvalue with respect to $M_x$ on the $k_x=\pi/2$ plane;
this is in analogy with the results of Ref. \onlinecite{tci_qpi_theory} for Pb$_{1-x}$Sn$_x$Te.
The second one is towards the $\bar Y$ cone ($\bq_3$);
in some cases we also observe a weak signal coming from the scattering to the $\bar\Gamma$ cone.
We stress, however, that the intensities (both absolute and relative) of these new QPI peaks depend on the exact choice of model parameters, as well as on the parameters used to simulate the tunneling process from the tip to the surface. We also note that these peaks arise primarily from the spectral weight in the first layer, which is mostly of $d$ character. Since $d$ states contribute a small fraction of the total weight of the surface states, their QPI peaks may -- depending on details of the tunneling process -- be masked by an incoherent background arising from $f$ states. This is illustrated in Fig.~\ref{fig_qpi}(i-l) where we have included a contribution from the second layer (which does carry $f$ weight); as a result, most QPI peaks dramatically lose contrast.
Even under these circumstances it is still possible to observe a weak signal from the $\bq_1$ channel.
%which could correspond to that seen in a recent QPI experiment \cite{hoffman_private_comm}

\begin{figure}[tb]
\includegraphics[width=0.49\textwidth]{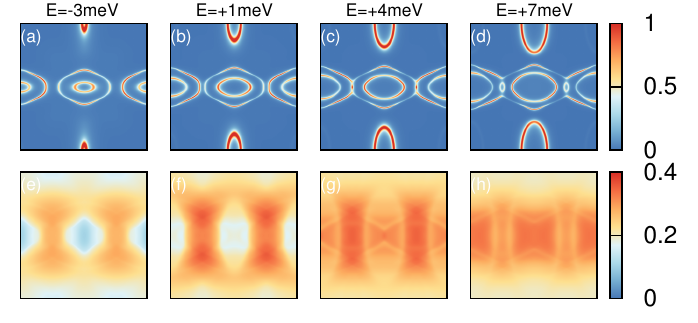}
\caption{Same as Fig.~\ref{fig_qpi}(a-h), but for set of parameters with a reduced ratio between hybridization and $f$ kinetic energy, see text. This case yield no structured QPI signal.
}\label{fig_qpi_2}
\end{figure}

\subsection{Numerical results: Other cases}
\label{sec:num_other}

We have repeated the same calculations for various other sets of parameters, and we describe a few representative cases in what follows.

\textit{Kondo breakdown.}
To model Kondo breakdown in the top layer, we take $\Delta E_f=100$\,eV, much larger than all other energy scales. Combined with the same $\Delta E_d=-0.5$\,eV, we obtained numerical data (not shown) that are almost indistinguishable from the example shown before in Fig.~\ref{fig_qpi}. We remark that a finite $\Delta E_d$ is needed in our set of parameters to increase $k_F$ and allow for the Lifshitz transition, but is not be needed for all sets of parameters (see below).

\textit{Reduced hybridization vs. kinetic energy.}
If we reduce the hybridization with respect to the $f$ kinetic energy, while keeping approximatively the same gap
($t^f\rightarrow 1.9 t^f$, $v\rightarrow 0.5 v$),
the QPI signal strength is strongly reduced, Fig.~\ref{fig_qpi_2}.
In particular, for energies below the Lifshitz transition, we observe no sign of nesting.
This can be reasonably linked to the slightly increased penetration length of surface states, which makes reconstruction a weaker perturbation.
In general, the appearance of QPI peaks and in particular their strength is found to be parameter-dependent.

\begin{figure}[tb]
\includegraphics[width=0.49\textwidth]{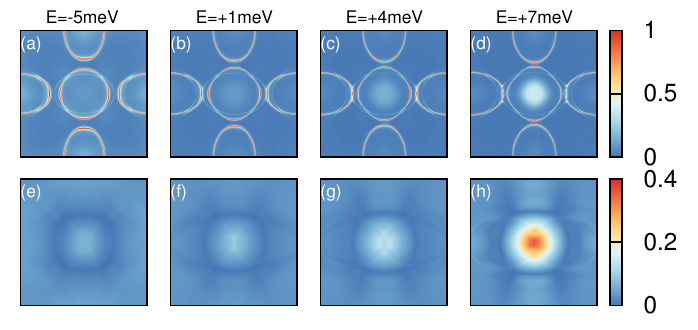}
\caption{Same as Fig.~\ref{fig_qpi}(a-h), but for
%no reconstruction potential $V_{rec}=0$, $\Delta E=-0.3$\,eV,
%and for
a weak reconstruction potential $\Vrec=0.2$\,eV, $\Delta E=-0.4$\,eV.
%, and $V_{rec}=0.5$\,eV, $\Delta E=-0.3$\,eV:
%in this second and third cases
Weak replicas of the cones can be observed in the isoenergy contour (a-d),
and no appreciable QPI signal is present (e-h), even though a signal corresponding to $\bq_2$ in Fig.~\ref{fig_qpi} starts to be visible.
%Color-scale units are arbitrary, but consistent among the three cases;
%in each case we fix $\Delta E$ in such a way to have approximatively always the same dispersion.
}\label{fig_qpi_small}
\end{figure}

\begin{figure}[tb]
\includegraphics[width=0.49\textwidth]{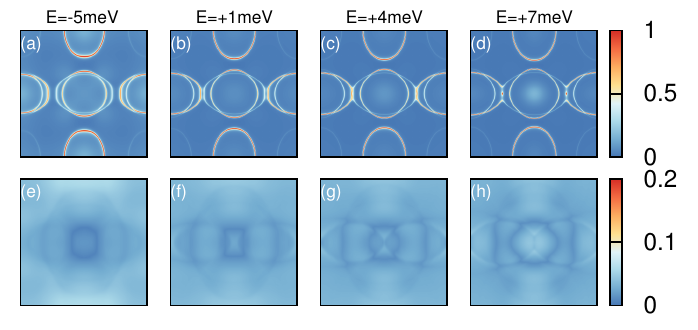}
\caption{Same as Fig.~\ref{fig_qpi}(a-h), but for
%no reconstruction potential $V_{rec}=0$, $\Delta E=-0.3$\,eV,
for a moderate reconstruction potential $\Vrec=0.5$\,eV, $\Delta E=-0.3$\,eV.
}\label{fig_qpi_small2}
\end{figure}

\textit{Surface scattering potential $\Delta E_d=0$.}
For a similar set of parameters (see Appendix) we are able to get a $k_F$ close to the experimental value and realize a Lifshitz transition with $\Delta E_f=100$\,eV
and $\Delta E_d=0$. The resulting QPI signal (not shown) is similar to Fig.~\ref{fig_qpi_2}.

\textit{Small reconstruction potential.}
For smaller reconstruction potential $\Vrec=0.2$\,eV and $\Vrec=0.5$\,eV, Figs. \ref{fig_qpi_small} and \ref{fig_qpi_small2}, the ARPES signal displays weak replicas of the cones appearing translated at the reconstruction wavevector $\bQ=(\pi,0)$.
In this case the QPI signal is very weak and almost indistinguishable from the case $\Vrec=0$, which, as remarked, shows no peaks.

%%%%%%%%%%%%%%%%%%%%%%%%%%%%%%%%%%%%%%%%%%%%%%%%%%%%%%%%%%%%%%%%%%%%%%%
%%%%%%%%%%%%%%%%%%%%%%%%%%%%%%%%%%%%%%%%%%%%%%%%%%%%%%%%%%%%%%%%%%%%%%%
%%%%%%%%%%%%%%%%%%%%%%%%%%%%%%%%%%%%%%%%%%%%%%%%%%%%%%%%%%%%%%%%%%%%%%%

\section{Conclusions}
\label{sec_concl}

In this paper we have studied the role of a periodic $2\times 1$ surface reconstruction on the topological surface states of {\sm} using a simple tight-binding model.
We qualitatively distinguish two cases, according to whether the effective reconstruction potential (ERP) acting on the surface states is weak or strong.
Weak ERP only produces a backfolding of surface bands, such that weak replicas of the original Dirac cones appear shifted by the reconstruction wavevector $\bQ=(\pi,0)$.
In contrast, strong ERP induces a particular crossing of Dirac cones, accompanied by a Lifshitz transition of in-gap states and the formation of a new Dirac cone, protected by mirror symmetry, at the edge of the small BZ. We have provided a numerical example for the case of strong ERP, and shown that new QPI peaks can appear as a consequence of this transition.

The ERP itself depends mostly on the weight of surface states on the first atomic layer(s), which in turn depends on microscopic details such as the penetration depth $\lambda$ of surface states, or the presence of a surface scattering potential. Our tight-binding approach cannot fully predict the strength of the ERP because detailed information on the structure of the reconstructed surface would be required. Ab-initio calculations could help, but to our knowledge no systematic studies have been conducted on reconstructed {\sm} surfaces.
Experimentally, existing ARPES data -- showing weak replicas only -- suggest a weak ERP which, however, is difficult to reconcile with our analysis: Using reasonable model assumptions, we tend to generically obtain strong ERP.
Possibly, the ERP is stronger than what appears from ARPES experiments, for example if the probed surface area contains large non-reconstructed (or disordered) regions which in turn do not contribute to band backfolding. We recall that some other correlated materials with strong periodic modulations also fail to display strong bandfolding effects in ARPES, presumably due to quenched disorder, one example being La$_{1.8-x}$Eu$_{0.2}$Sr$_x$CuO$_4$ \cite{zabol09}.

Based on our results, we suggest that small-spot ARPES on $2\times1$ reconstructed surfaces as well as careful QPI studies, searching for additional peaks appearing above the Lifshitz transition,
could clarify and further elucidate the surface-reconstruction effects in {\sm}.

\acknowledgments

We thank J. E. Hoffman, J. Denlinger, H. Fehske, L. Fritz, D. K. Morr, H. Pirie, O. Rader, and S. Wirth for discussions and collaborations on related work, and J. E. Hoffman for sharing experimental data.
This research was supported by the DFG through SFB 1143 and GRK 1621
as well as by the Helmholtz association through VI-521.

%%%%%%%%%%%%%%%%%%%%%%%%%%%%%%%%%%%%%%%%%%%%%%%%%%%%%%%%%%%%%%%%%%%%%%%
%%%%%%%%%%%%%%%%%%%%%%%%%%%%%%%%%%%%%%%%%%%%%%%%%%%%%%%%%%%%%%%%%%%%%%%
%%%%%%%%%%%%%%%%%%%%%%%%%%%%%%%%%%%%%%%%%%%%%%%%%%%%%%%%%%%%%%%%%%%%%%%

\appendix

\section{Tight-binding modeling}
%{\sm} crystallizes in the simple cubic ($sc$) structure, with a $sc$ 3D BZ. %(see Fig.~\ref{fig_3dbz} (a)-(b)).
%For simplicity, in what follows, we set the lattice spacing $a=4.13$\AA$=1$.

\subsection{Bulk model}

To model {\sm}, we only retain Sm atoms, and we build a tight-binding model out of the $d_{x^2-y^2}$ - $d_{z^2}$ quadruplet in the $d$ shell,
and of the $\Gamma_8$ quadruplet in the $f$ shell, where ${|\Gamma_8^{(1)}\pm\rangle} = {\sqrt\frac{5}{6}|\pm\frac{5}{2}\rangle + \sqrt\frac{1}{6}|\mp\frac{3}{2}\rangle}$,
${|\Gamma_8^{(2)}\pm\rangle} = {|\pm\frac{1}{2}\rangle}$\cite{takimoto,tki_cubic,prbr_io_smb6,smb6_tci_io}.
We use two kinetic energy terms for $d$ states
($t_1^d=\langle d_{z^2}|H|d_{z^2}\rangle_{001}$,  $t_2^d=\langle d_{z^2}|H|d_{z^2}\rangle_{110}$, where the suffix $001$ or $110$ denotes the direction along which the matrix element is considered),
two kinetic energy terms for $f$ states
($t_1^f=\langle \Gamma_8^{(2)}+|H|\Gamma_8^{(2)}+\rangle_{001}$,  $t_2^f=\langle \Gamma_8^{(2)}+|H|\Gamma_8^{(2)}+\rangle_{110}$),
two hybridization terms
($v_1=\langle d_{z^2}\uparrow|H|\Gamma_8^{(2)}+\rangle_{001}$,  $v_2=\langle d_{z^2}\uparrow|H|\Gamma_8^{(2)}+\rangle_{110}$),
and one on-site energy difference between $d$ and $f$ states $\epsilon_d-\epsilon_f$.
The kinetic energy is fixed to give at the three $X=(0,0,\pi)=(0,\pi,0)=(\pi,0,0)$ points a minimum in the $d$ shell, with $d_{x^2-y^2}$ symmetry at $(0,0,\pi)$, and a maximum in the $f$ shell;
this leads to band inversion at the $X$ points, and to the topological $\mathbb{Z}_2$ indices $(\nu_0,\nu_1,\nu_2,\nu_3)=(1,1,1,1)$.
The hybridization is chosen to be in the $\C^+_{k_z=0}=+2$, $\C^+_{k_z=\pi}=+1$, $\C^+_{k_x=k_y}=-1$ phase, where $\C^+$ are mirror Chern numbers\cite{smb6_tci,smb6_tci_io,smb6_tci_legner}:
\be\label{chern_number}
\C^+_{\overline{BZ}}=\frac{\ii}{2\pi}\sum_{a,b=1}^2\epsilon_{ab}\sum_{n=1}^N\int_{\overline{BZ}}d^2\bk \langle \partial_a u_n^+(\bk)| \partial_b u_n^+(\bk)\rangle,
\ee
where $M|u_n^+(\bk)\rangle=+ \ii|u_n^+(\bk)\rangle$ and $\bk$ lying in the plane $\overline{BZ}$  which is invariant under the symmetry operator $M$
($M$=$M_z$ when $\overline{BZ}$ is $k_z=0$ or $k_z=\pi$, $M=M_{x-y}$ when $\overline{BZ}$ is $k_x=k_y$).
This leads to a positive winding number on $\bar X$ cones\cite{smb6_tci_io} as observed experimentally\cite{smb6_arpes_mesot_spin},
and is relevant for the QPI signal and the spin structure of surface states, but not for their dispersion.

The Hamiltonian
\be
H=H_{kin}+H_{hybr}
\ee
is composed by the kinetic energy term:
\bea
H_{kin}=
\left( \begin{array}{ll}
H_d & 0 \\
0 &H_f
\end{array}\right),
\eea
and by the hybridization:
\bea
H_{hybr}=
\left( \begin{array}{ll}
0 & \ii H_{df} \\
-\ii H_{df}^\dagger &0
\end{array}\right).
\eea
The kinetic energy $H_i$, $i=d/f$, is  diagonal in the (pseudo)spin index, with basis $d_{x^2-y^2}$ and $d_{z^2}$ when $i=d$,
and $\Gamma_8^{(1)}$, $\Gamma_8^{(2)}$ when $i=f$;
the hybridization $H_{df}$ is  nondiagonal in the (pseudo)spin index, with basis $d_{x^2-y^2}\uparrow$, $d_{x^2-y^2}\downarrow$, $d_{z^2}\uparrow$ and $d_{z^2}\downarrow$ for rows,
and  $\Gamma_8^{(1)}+$, $\Gamma_8^{(1)}-$, $\Gamma_8^{(2)}+$ $\Gamma_8^{(2)}-$ for columns.

Their explicit form is as follows ($c_x=\cos k_x$, $c_y=\cos k_y$, $c_z=\cos k_z$, $s_x=\sin k_x$, $s_y=\sin k_y$, $s_z=\sin k_z$, $s_\pm=s_x\pm \ii s_y$, $cs_\pm=c_ys_x\pm \ii c_x s_y$):
\begin{widetext}
%\onecolumngrid
\bea
H_{i}=
\left( \begin{array}{ll}
\epsilon^i+3(c_x+c_y)\left({t_1^i}/{2}+c_zt_2^i\right)& \sqrt{3}(c_x-c_y)\left(-{t_1^i}/{2}+c_zt_2^i\right)\\
\sqrt{3}(c_x-c_y)\left(-{t_1^i}/{2}+c_zt_2^i\right)& \epsilon^i+(c_x+c_y)\left({t_1^i}/{2}+c_zt_2^i\right)+2c_zt_1^i+4c_xc_yt_2^i
\end{array}\right),\hspace{10pt}i=d/f,
\eea
\bea
H_{df}=
\left( \begin{array}{llll}
3v_2(c_x+c_y)s_z & 3s_-(v_1/2+c_zv_2) & \sqrt{3}(c_x-c_y)s_zv_2&\sqrt{3}s_+(-v_1/2+c_zv_2)\\
3s_+(v_1/2+c_zv_2) & -3v_2(c_x+c_y)s_z & \sqrt{3}s_-(-v_1/2+c_zv_2)&-\sqrt{3}(c_x-c_y)s_zv_2\\
\sqrt{3}(c_x-c_y)s_zv_2 & \sqrt{3}s_+(-v_1/2+c_zv_2) & s_z[2v_1+(c_x+c_y)v_2] & s_-(v_1/2+c_zv_2)+4v_2cs_-\\
\sqrt{3}s_-(-v_1/2+c_zv_2)& -\sqrt{3}(c_x-c_y)s_zv_2 &s_+(v_1/2+c_zv_2)+4v_2cs_+&-s_z[2v_1+(c_x+c_y)v_2]
\end{array}\right).
\eea
\end{widetext}
%\twocolumngrid
%v_1=\eta_z^{v1}
%v_2=\eta_z^{v2}

%For the slab calculation, we introduce a surface scattering potential $\Delta E$ which modifies the on-site energy of states on the first layer,
%and is fixed in such a way to reproduce as closely as possible the experimental ARPES dispersion of surface states.

\begin{figure}[h]
\includegraphics[width=0.48\textwidth]{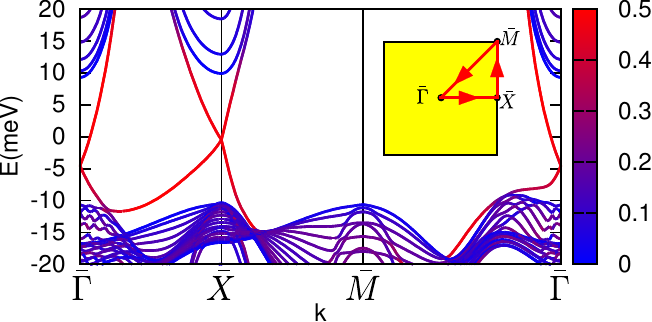}
\caption{
Bandstructure of the {\sm} model employed in the paper, in a 15-layer slab geometry without surface reconstruction, plotted along the $\bar \Gamma \bar X \bar M \bar \Gamma$ path in the large surface BZ.
}\label{fig_bands2}
\end{figure}

Parameters are taken from the tight-binding calculations of Refs. \onlinecite{pub6,prbr_io_smb6}, with a slave-boson-like renormalization of the $f$ kinetic energy by $b^2$
and of the hybridization by $b$, with $b^2\sim 0.08$,
and fine tuned to better match ARPES experimental data.
For Figures \ref{fig_bands}, \ref{fig_dos} and \ref{fig_qpi} we increased the value of the hybridization by roughly $50\%$ with respect to this estimation to better highlight QPI features; this however does not affect appreciably the ERP, which would be in any case large.
%however, to roughly recover the experimental ARPES value of the bulk gap, we increased the hybridization by roughly $50\%$, with the net result of a reduced penetration length of surface states
%and an increased effect of the reconstruction.
In particular we employ:
$\epsilon^d-\epsilon^f=1.8$\,eV,
$t^d_1=-0.76$\,eV,		%0.8*(0.8)
$t^d_2=0.2$\,eV,	%0.8*(-0.3)
$t^f_1=3.2$\,meV,	%0.003*(4)
$t^f_2=-1.6$\,meV,	%0.003*(-2)
$v_1=-84$\,meV,	%0.06*(-2.1)
$v_2=24$\,meV,		%0.06(0.6)
which lead to a bulk gap $\Delta=18$\,meV, to $\epsilon^f=-10$\,meV with respect to the chemical potential, and to a level occupation $n_d=0.51$, $n_f=3.49$,
which corresponds to a Sm$^{2.5+}$ valence;
% corresponding to a valence Sm$^{2.51+}$.
the chemical potential is set in the middle of the bulk gap.
In Fig.~\ref{fig_bands2} we show the dispersion for a slab without reconstruction;
the same data is shown in Fig.~\ref{fig_bands}(a) in the small BZ.

\subsection{Surface reconstruction}

We model reconstruction by applying a large scattering potential $\Vrec=100$\,eV on sites every second row of the top layer; we neglect any relaxation effect that the removal of these top atoms can cause. We also apply a surface scattering potential $\Delta E=\Delta E_d=\Delta E_f=-0.5$\,eV,
which  is set to align the energies of the Dirac cones to approximatively reproduce a situation close to the experimental one, allowing for the Lifshitz transition of surface states.
The penetration length $\lambda$ at the Fermi energy is about two atomic layers for all cones.

In the Kondo-breakdown scenario we set $\Delta E_f=100$\,eV for $f$ states, but still retain a finite $\Delta E_d=-0.5$\,eV for $d$ states; in this case no appreciable modification of the dispersion or of the QPI signal is observed with respect to Figures \ref{fig_bands} and \ref{fig_qpi} of the main text.

In Fig.~\ref{fig_qpi_2} we use $t^f_1=6$\,meV, $t^f_2=-3$meV, $v_1=-42$\,meV, $v_2=12$\,meV, $\Delta E=0.1$\,eV, and keep the other parameters fixed.

We are able to achieve the Lifshitz transition with $\Delta E_d=0$, when $t^f_1=4.8$\,meV, $t^f_2=-1.8$\,meV, $v_1=-46$\,meV, $v_2=13$\,meV, and the other parameters fixed. For this parameter set we do not show results, but QPI patterns are very similar to those of Fig.~\ref{fig_qpi_2}.

\subsection{QPI}

QPI figures are generated with a $400\times 400$ mesh on a 15-layer slab, using the scattering matrix technique (see Ref. \onlinecite{prb_io_tki} for technical details)
with an artificial broadening of $\delta=1$\,meV (Figs.~\ref{fig_qpi_small} and~\ref{fig_qpi_small2}: $\delta=2$\,meV).
We take a $30$\,meV scatterer in both $d$ and the $f$ channels in the first two layers in sublattice $a$,
but results were found to be qualitatively independent of the choice of the scattering potential.
As remarked earlier, quantitative results depend on the exact choice of the parameters \cite{renorm_foot}.

In the QPI figures, we only report the Fourier transform of the impurity-induced piece of the LDOS, i.e., structural peaks at momenta $(0,0)$ and $(\pi,0)$ (the reconstruction wavevector) are not shown.

\bibliographystyle{apsrev4-1}
\bibliography{tki}

%Merlin.mbs v4.21 2009-07-09.
\begin{thebibliography}{10}%
\makeatletter
\providecommand \@ifxundefined [1]{%
 \ifx #1\undefined \expandafter \@firstoftwo
 \else \expandafter \@secondoftwo
\fi
}%
\providecommand \@ifnum [1]{%
 \ifnum #1\expandafter \@firstoftwo
 \else \expandafter \@secondoftwo
\fi
}%
\providecommand \enquote [1]{``#1''}%
\providecommand \bibnamefont  [1]{#1}%
\providecommand \bibfnamefont [1]{#1}%
\providecommand \citenamefont [1]{#1}%
\providecommand\href[0]{\@sanitize\@href}%
\providecommand\@href[1]{\endgroup\@@startlink{#1}\endgroup\@@href}%
\providecommand\@@href[1]{#1\@@endlink}%
\providecommand \@sanitize [0]{\begingroup\catcode`\&12\catcode`\#12\relax}%
\@ifxundefined \pdfoutput {\@firstoftwo}{%
 \@ifnum{\z@=\pdfoutput}{\@firstoftwo}{\@secondoftwo}%
}{%
 \providecommand\@@startlink[1]{\leavevmode\special{html:<a href="#1">}}%
 \providecommand\@@endlink[0]{\special{html:</a>}}%
}{%
 \providecommand\@@startlink[1]{%
  \leavevmode
  \pdfstartlink
   attr{/Border[0 0 1 ]/H/I/C[0 1 1]}%
   user{/Subtype/Link/A<</Type/Action/S/URI/URI(#1)>>}%
  \relax
 }%
 \providecommand\@@endlink[0]{\pdfendlink}%
}%
\providecommand \url  [0]{\begingroup\@sanitize \@url }%
\providecommand \@url [1]{\endgroup\@href {#1}{\urlprefix}}%
\providecommand \urlprefix [0]{URL }%
\providecommand \Eprint[0]{\href }%
\@ifxundefined \urlstyle {%
  \providecommand \doi [1]{doi:\discretionary{}{}{}#1}%
}{%
  \providecommand \doi [0]{doi:\discretionary{}{}{}\begingroup
  \urlstyle{rm}\Url }%
}%
\providecommand \doibase [0]{http://dx.doi.org/}%
\providecommand \Doi[1]{\href{\doibase#1}}%
\providecommand \bibAnnote [3]{%
  \BibitemShut{#1}%
  \begin{quotation}\noindent
    \textsc{Key:}\ #2\\\textsc{Annotation:}\ #3%
  \end{quotation}%
}%
\providecommand \bibAnnoteFile [2]{%
  \IfFileExists{#2}{\bibAnnote {#1} {#2} {\input{#2}}}{}%
}%
\providecommand \typeout [0]{\immediate \write \m@ne }%
\providecommand \selectlanguage [0]{\@gobble}%
\providecommand \bibinfo [0]{\@secondoftwo}%
\providecommand \bibfield [0]{\@secondoftwo}%
\providecommand \translation [1]{[#1]}%
\providecommand \BibitemOpen[0]{}%
\providecommand \bibitemStop [0]{}%
\providecommand \bibitemNoStop [0]{.\EOS\space}%
\providecommand \EOS [0]{\spacefactor3000\relax}%
\providecommand \BibitemShut [1]{\csname bibitem#1\endcsname}%
%</preamble>
\bibitem{takimoto}%
  \BibitemOpen
  \bibfield{author}{%
  \bibinfo {author} {\bibfnamefont{T.}~\bibnamefont{Takimoto}},\ }%
  \bibfield{journal}{%
  \bibinfo {journal} {J. Phys. Soc. Jpn.}\ }%
  \textbf{\bibinfo {volume} {80}},\ \bibinfo {pages} {123710} (\bibinfo {year}
  {2011})%
  \bibAnnoteFile{NoStop}{takimoto}%
\bibitem{lu_smb6_gutz}%
  \BibitemOpen
  \bibfield{author}{%
  \bibinfo {author} {\bibfnamefont{F.}~\bibnamefont{Lu}}, \bibinfo {author}
  {\bibfnamefont{J.}~\bibnamefont{Zhao}}, \bibinfo {author}
  {\bibfnamefont{H.}~\bibnamefont{Weng}}, \bibinfo {author}
  {\bibfnamefont{Z.}~\bibnamefont{Fang}},\ and\ \bibinfo {author}
  {\bibfnamefont{X.}~\bibnamefont{Dai}},\ }%
  \bibfield{journal}{%
  \bibinfo {journal} {Phys. Rev. Lett.}\ }%
  \textbf{\bibinfo {volume} {110}},\ \bibinfo {pages} {096401} (\bibinfo {year}
  {2013})%
  \bibAnnoteFile{NoStop}{lu_smb6_gutz}%
\bibitem{tki_cubic}%
  \BibitemOpen
  \bibfield{author}{%
  \bibinfo {author} {\bibfnamefont{V.}~\bibnamefont{Alexandrov}}, \bibinfo
  {author} {\bibfnamefont{M.}~\bibnamefont{Dzero}},\ and\ \bibinfo {author}
  {\bibfnamefont{P.}~\bibnamefont{Coleman}},\ }%
  \bibfield{journal}{%
  \bibinfo {journal} {Phys. Rev. Lett.}\ }%
  \textbf{\bibinfo {volume} {111}},\ \bibinfo {pages} {226403} (\bibinfo {year}
  {2013})%
  \bibAnnoteFile{NoStop}{tki_cubic}%
\bibitem{tki1}%
  \BibitemOpen
  \bibfield{author}{%
  \bibinfo {author} {\bibfnamefont{M.}~\bibnamefont{Dzero}}, \bibinfo {author}
  {\bibfnamefont{K.}~\bibnamefont{Sun}}, \bibinfo {author}
  {\bibfnamefont{V.}~\bibnamefont{Galitski}},\ and\ \bibinfo {author}
  {\bibfnamefont{P.}~\bibnamefont{Coleman}},\ }%
  \bibfield{journal}{%
  \bibinfo {journal} {Phys. Rev. Lett.}\ }%
  \textbf{\bibinfo {volume} {104}},\ \bibinfo {pages} {106408} (\bibinfo {year}
  {2010})%
  \bibAnnoteFile{NoStop}{tki1}%
\bibitem{smb6_tci}%
  \BibitemOpen
  \bibfield{author}{%
  \bibinfo {author} {\bibfnamefont{M.}~\bibnamefont{Ye}}, \bibinfo {author}
  {\bibfnamefont{J.~W.}\ \bibnamefont{Allen}},\ and\ \bibinfo {author}
  {\bibfnamefont{K.}~\bibnamefont{Sun}},\ }%
  \bibinfo {journal} {preprint arXiv:1307.7191}%
  \bibAnnoteFile{NoStop}{smb6_tci}%
\bibitem{smb6_tci_io}%
  \BibitemOpen
\bibfield{journal}{%
    }%
  \bibfield{author}{%
  \bibinfo {author} {\bibfnamefont{P.~P.}\ \bibnamefont{Baruselli}}\ and\
  \bibinfo {author} {\bibfnamefont{M.}~\bibnamefont{Vojta}},\ }%
  \bibinfo {journal} {preprint arXiv:1505.03507}%
  \bibAnnoteFile{NoStop}{smb6_tci_io}%
\bibitem{smb6_tci_legner}%
  \BibitemOpen
\bibfield{journal}{%
    }%
  \bibfield{author}{%
  \bibinfo {author} {\bibfnamefont{M.}~\bibnamefont{Legner}}, \bibinfo {author}
  {\bibfnamefont{A.}~\bibnamefont{R\"uegg}},\ and\ \bibinfo {author}
  {\bibfnamefont{M.}~\bibnamefont{Sigrist}},\ }%
  \bibinfo {journal} {preprint arXiv:1505.02987}%
  \bibAnnoteFile{NoStop}{smb6_tci_legner}%
\bibitem{wolgast_smb6}%
  \BibitemOpen
\bibfield{journal}{%
    }%
  \bibfield{author}{%
  \bibinfo {author} {\bibfnamefont{S.}~\bibnamefont{Wolgast}}, \bibinfo
  {author} {\bibfnamefont{C.}~\bibnamefont{Kurdak}}, \bibinfo {author}
  {\bibfnamefont{K.}~\bibnamefont{Sun}}, \bibinfo {author}
  {\bibfnamefont{J.~W.}\ \bibnamefont{Allen}}, \bibinfo {author}
  {\bibfnamefont{D.-J.}\ \bibnamefont{Kim}},\ and\ \bibinfo {author}
  {\bibfnamefont{Z.}~\bibnamefont{Fisk}},\ }%
  \bibfield{journal}{%
  \bibinfo {journal} {Phys. Rev. B}\ }%
  \textbf{\bibinfo {volume} {88}},\ \bibinfo {pages} {180405(R)} (\bibinfo
  {year} {2013})%
  \bibAnnoteFile{NoStop}{wolgast_smb6}%
\bibitem{fisk_smb6_topss}%
  \BibitemOpen
  \bibfield{author}{%
  \bibinfo {author} {\bibfnamefont{D.~J.}\ \bibnamefont{Kim}}, \bibinfo
  {author} {\bibfnamefont{J.}~\bibnamefont{Xia}},\ and\ \bibinfo {author}
  {\bibfnamefont{Z.}~\bibnamefont{Fisk}},\ }%
  \bibfield{journal}{%
  \bibinfo {journal} {Nature Mat.}\ }%
  \textbf{\bibinfo {volume} {13}},\ \bibinfo {pages} {466} (\bibinfo {year}
  {2014})%
  \bibAnnoteFile{NoStop}{fisk_smb6_topss}%
\bibitem{smb6_junction_prx}%
  \BibitemOpen
  \bibfield{author}{%
  \bibinfo {author} {\bibfnamefont{X.}~\bibnamefont{Zhang}}, \bibinfo {author}
  {\bibfnamefont{N.~P.}\ \bibnamefont{Butch}}, \bibinfo {author}
  {\bibfnamefont{P.}~\bibnamefont{Syers}}, \bibinfo {author}
  {\bibfnamefont{S.}~\bibnamefont{Ziemak}}, \bibinfo {author}
  {\bibfnamefont{R.~L.}\ \bibnamefont{Greene}},\ and\ \bibinfo {author}
  {\bibfnamefont{J.}~\bibnamefont{Paglione}},\ }%
  \bibfield{journal}{%
  \bibinfo {journal} {Phys. Rev. X}\ }%
  \textbf{\bibinfo {volume} {3}},\ \bibinfo {pages} {011011} (\bibinfo {year}
  {2013})%
  \bibAnnoteFile{NoStop}{smb6_junction_prx}%
\bibitem{lixiang_smb6}%
  \BibitemOpen
  \bibfield{author}{%
  \bibinfo {author} {\bibfnamefont{G.}~\bibnamefont{Li}}, \bibinfo {author}
  {\bibfnamefont{Z.}~\bibnamefont{Xiang}}, \bibinfo {author}
  {\bibfnamefont{F.}~\bibnamefont{Yu}}, \bibinfo {author}
  {\bibfnamefont{T.}~\bibnamefont{Asaba}}, \bibinfo {author}
  {\bibfnamefont{B.}~\bibnamefont{Lawson}}, \bibinfo {author}
  {\bibfnamefont{P.}~\bibnamefont{Cai}}, \bibinfo {author}
  {\bibfnamefont{C.}~\bibnamefont{Tinsman}}, \bibinfo {author}
  {\bibfnamefont{A.}~\bibnamefont{Berkley}}, \bibinfo {author}
  {\bibfnamefont{S.}~\bibnamefont{Wolgast}}, \bibinfo {author}
  {\bibfnamefont{Y.~S.}\ \bibnamefont{Eo}}, \bibinfo {author}
  {\bibfnamefont{D.-J.}\ \bibnamefont{Kim}}, \bibinfo {author}
  {\bibfnamefont{C.}~\bibnamefont{Kurdak}}, \bibinfo {author}
  {\bibnamefont{Allen}}, \bibinfo {author}
  {\bibfnamefont{K.}~\bibnamefont{J.~W.~Sun}}, \bibinfo {author}
  {\bibfnamefont{X.~H.}\ \bibnamefont{Chen}}, \bibinfo {author}
  {\bibfnamefont{Y.~Y.}\ \bibnamefont{Wang}}, \bibinfo {author}
  {\bibfnamefont{Z.}~\bibnamefont{Fisk}},\ and\ \bibinfo {author}
  {\bibfnamefont{L.}~\bibnamefont{Li}},\ }%
  \bibinfo {journal} {preprint arXiv:1306.5221}%
  \bibAnnoteFile{NoStop}{lixiang_smb6}%
\bibitem{neupane_smb6}%
  \BibitemOpen
\bibfield{journal}{%
    }%
  \bibfield{author}{%
  \bibinfo {author} {\bibfnamefont{M.}~\bibnamefont{Neupane}}, \bibinfo
  {author} {\bibfnamefont{N.}~\bibnamefont{Alidoust}}, \bibinfo {author}
  {\bibfnamefont{S.}~\bibnamefont{Xu}}, \bibinfo {author}
  {\bibfnamefont{T.}~\bibnamefont{Kondo}}, \bibinfo {author}
  {\bibfnamefont{Y.}~\bibnamefont{Ishida}}, \bibinfo {author}
  {\bibfnamefont{D.-J.}\ \bibnamefont{Kim}}, \bibinfo {author}
  {\bibfnamefont{C.}~\bibnamefont{Liu}}, \bibinfo {author}
  {\bibfnamefont{I.}~\bibnamefont{Belopolski}}, \bibinfo {author}
  {\bibfnamefont{Y.}~\bibnamefont{Jo}}, \bibinfo {author}
  {\bibfnamefont{T.-R.}\ \bibnamefont{Chang}}, \bibinfo {author}
  {\bibfnamefont{H.-T.}\ \bibnamefont{Jeng}}, \bibinfo {author}
  {\bibfnamefont{T.}~\bibnamefont{Durakiewicz}}, \bibinfo {author}
  {\bibfnamefont{L.}~\bibnamefont{Balicas}}, \bibinfo {author}
  {\bibfnamefont{H.}~\bibnamefont{Lin}}, \bibinfo {author}
  {\bibfnamefont{A.}~\bibnamefont{Bansil}}, \bibinfo {author}
  {\bibfnamefont{S.}~\bibnamefont{Shin}}, \bibinfo {author}
  {\bibfnamefont{Z.}~\bibnamefont{Fisk}},\ and\ \bibinfo {author}
  {\bibfnamefont{M.~Z.}\ \bibnamefont{Hasan}},\ }%
  \bibfield{journal}{%
  \bibinfo {journal} {Nature Comm.}\ }%
  \textbf{\bibinfo {volume} {4}},\ \bibinfo {pages} {2991} (\bibinfo {year}
  {2013})%
  \bibAnnoteFile{NoStop}{neupane_smb6}%
\bibitem{mesot_smb6}%
  \BibitemOpen
  \bibfield{author}{%
  \bibinfo {author} {\bibfnamefont{N.}~\bibnamefont{Xu}}, \bibinfo {author}
  {\bibfnamefont{X.}~\bibnamefont{Shi}}, \bibinfo {author}
  {\bibfnamefont{P.~K.}\ \bibnamefont{Biswas}}, \bibinfo {author}
  {\bibfnamefont{C.~E.}\ \bibnamefont{Matt}}, \bibinfo {author}
  {\bibfnamefont{R.~S.}\ \bibnamefont{Dhaka}}, \bibinfo {author}
  {\bibfnamefont{Y.}~\bibnamefont{Huang}}, \bibinfo {author}
  {\bibfnamefont{N.~C.}\ \bibnamefont{Plumb}}, \bibinfo {author}
  {\bibfnamefont{M.}~\bibnamefont{Radovic}}, \bibinfo {author}
  {\bibfnamefont{J.~H.}\ \bibnamefont{Dil}}, \bibinfo {author}
  {\bibfnamefont{E.}~\bibnamefont{Pomjakushina}}, \bibinfo {author}
  {\bibfnamefont{K.}~\bibnamefont{Conder}}, \bibinfo {author}
  {\bibfnamefont{A.}~\bibnamefont{Amato}}, \bibinfo {author}
  {\bibfnamefont{Z.}~\bibnamefont{Salman}}, \bibinfo {author}
  {\bibfnamefont{D.~M.}\ \bibnamefont{Paul}}, \bibinfo {author}
  {\bibfnamefont{J.}~\bibnamefont{Mesot}}, \bibinfo {author}
  {\bibfnamefont{H.}~\bibnamefont{Ding}},\ and\ \bibinfo {author}
  {\bibfnamefont{M.}~\bibnamefont{Shi}},\ }%
  \bibfield{journal}{%
  \bibinfo {journal} {Phys. Rev. B}\ }%
  \textbf{\bibinfo {volume} {88}},\ \bibinfo {pages} {121102} (\bibinfo {year}
  {2013})%
  \bibAnnoteFile{NoStop}{mesot_smb6}%
\bibitem{smb6_arpes_feng}%
  \BibitemOpen
  \bibfield{author}{%
  \bibinfo {author} {\bibfnamefont{J.}~\bibnamefont{Jiang}}, \bibinfo {author}
  {\bibfnamefont{S.}~\bibnamefont{Li}}, \bibinfo {author}
  {\bibfnamefont{T.}~\bibnamefont{Zhang}}, \bibinfo {author}
  {\bibfnamefont{Z.}~\bibnamefont{Sun}}, \bibinfo {author}
  {\bibfnamefont{F.}~\bibnamefont{Chen}}, \bibinfo {author}
  {\bibfnamefont{Z.}~\bibnamefont{Ye}}, \bibinfo {author}
  {\bibfnamefont{M.}~\bibnamefont{Xu}}, \bibinfo {author}
  {\bibfnamefont{Q.}~\bibnamefont{Ge}}, \bibinfo {author}
  {\bibfnamefont{S.}~\bibnamefont{Tan}}, \bibinfo {author}
  {\bibfnamefont{X.}~\bibnamefont{Niu}}, \bibinfo {author}
  {\bibfnamefont{M.}~\bibnamefont{Xia}}, \bibinfo {author}
  {\bibfnamefont{B.}~\bibnamefont{Xie}}, \bibinfo {author}
  {\bibfnamefont{Y.}~\bibnamefont{Li}}, \bibinfo {author}
  {\bibfnamefont{X.}~\bibnamefont{Chen}}, \bibinfo {author}
  {\bibfnamefont{H.}~\bibnamefont{Wen}},\ and\ \bibinfo {author}
  {\bibfnamefont{D.}~\bibnamefont{Feng}},\ }%
  \bibfield{journal}{%
  \bibinfo {journal} {Nature Comm.}\ }%
  \textbf{\bibinfo {volume} {4}},\ \bibinfo {pages} {3010} (\bibinfo {year}
  {2013})%
  \bibAnnoteFile{NoStop}{smb6_arpes_feng}%
\bibitem{smb6_arpes_reinert}%
  \BibitemOpen
  \bibfield{author}{%
  \bibinfo {author} {\bibfnamefont{C.-H.}\ \bibnamefont{Min}}, \bibinfo
  {author} {\bibfnamefont{P.}~\bibnamefont{Lutz}}, \bibinfo {author}
  {\bibfnamefont{S.}~\bibnamefont{Fiedler}}, \bibinfo {author}
  {\bibfnamefont{B.}~\bibnamefont{Kang}}, \bibinfo {author}
  {\bibfnamefont{B.}~\bibnamefont{Cho}}, \bibinfo {author}
  {\bibfnamefont{H.-D.}\ \bibnamefont{Kim}}, \bibinfo {author}
  {\bibfnamefont{H.}~\bibnamefont{Bentmann}},\ and\ \bibinfo {author}
  {\bibfnamefont{F.}~\bibnamefont{Reinert}},\ }%
  \bibfield{journal}{%
  \bibinfo {journal} {Phys. Rev. Lett.}\ }%
  \textbf{\bibinfo {volume} {112}},\ \bibinfo {pages} {226402} (\bibinfo {year}
  {2014})%
  \bibAnnoteFile{NoStop}{smb6_arpes_reinert}%
\bibitem{smb6_past_allen}%
  \BibitemOpen
  \bibfield{author}{%
  \bibinfo {author} {\bibfnamefont{J.~D.}\ \bibnamefont{Denlinger}}, \bibinfo
  {author} {\bibfnamefont{J.~W.}\ \bibnamefont{Allen}}, \bibinfo {author}
  {\bibfnamefont{J.-S.}\ \bibnamefont{Kang}}, \bibinfo {author}
  {\bibfnamefont{K.}~\bibnamefont{Sun}}, \bibinfo {author}
  {\bibfnamefont{B.-I.}\ \bibnamefont{Min}}, \bibinfo {author}
  {\bibfnamefont{D.-J.}\ \bibnamefont{Kim}},\ and\ \bibinfo {author}
  {\bibfnamefont{Z.}~\bibnamefont{Fisk}},\ }%
  \bibinfo {journal} {preprint arXiv:1312.6636}%
  \bibAnnoteFile{NoStop}{smb6_past_allen}%
\bibitem{smb6_arpes_mesot_spin}%
  \BibitemOpen
\bibfield{journal}{%
    }%
  \bibfield{author}{%
  \bibinfo {author} {\bibfnamefont{N.}~\bibnamefont{Xu}}, \bibinfo {author}
  {\bibfnamefont{P.~K.}\ \bibnamefont{Biswas}}, \bibinfo {author}
  {\bibfnamefont{J.~H.}\ \bibnamefont{Dil}}, \bibinfo {author}
  {\bibfnamefont{R.~S.}\ \bibnamefont{Dhaka}}, \bibinfo {author}
  {\bibfnamefont{G.}~\bibnamefont{Landolt}}, \bibinfo {author}
  {\bibfnamefont{S.}~\bibnamefont{Muff}}, \bibinfo {author}
  {\bibfnamefont{C.~E.}\ \bibnamefont{Matt}}, \bibinfo {author}
  {\bibfnamefont{X.}~\bibnamefont{Shi}}, \bibinfo {author}
  {\bibfnamefont{N.~C.}\ \bibnamefont{Plumb}}, \bibinfo {author}
  {\bibfnamefont{M.}~\bibnamefont{Radovic}}, \bibinfo {author}
  {\bibfnamefont{E.}~\bibnamefont{Pomjakushina}}, \bibinfo {author}
  {\bibfnamefont{K.}~\bibnamefont{Conder}}, \bibinfo {author}
  {\bibfnamefont{A.}~\bibnamefont{Amato}}, \bibinfo {author}
  {\bibfnamefont{S.}~\bibnamefont{Borisenko}}, \bibinfo {author}
  {\bibfnamefont{R.}~\bibnamefont{Yu}}, \bibinfo {author}
  {\bibfnamefont{H.-M.}\ \bibnamefont{Weng}}, \bibinfo {author}
  {\bibfnamefont{Z.}~\bibnamefont{Fang}}, \bibinfo {author}
  {\bibfnamefont{X.}~\bibnamefont{Dai}}, \bibinfo {author}
  {\bibfnamefont{J.}~\bibnamefont{Mesot}}, \bibinfo {author}
  {\bibfnamefont{H.}~\bibnamefont{Ding}},\ and\ \bibinfo {author}
  {\bibfnamefont{M.}~\bibnamefont{Shi}},\ }%
  \bibfield{journal}{%
  \bibinfo {journal} {Nature Comm.}\ }%
  \textbf{\bibinfo {volume} {5}},\ \bibinfo {pages} {4566} (\bibinfo {year}
  {2014})%
  \bibAnnoteFile{NoStop}{smb6_arpes_mesot_spin}%
\bibitem{sawatzky_smb6}%
  \BibitemOpen
  \bibfield{author}{%
  \bibinfo {author} {\bibfnamefont{Z.-H.}\ \bibnamefont{Zhu}}, \bibinfo
  {author} {\bibfnamefont{A.}~\bibnamefont{Nicolaou}}, \bibinfo {author}
  {\bibfnamefont{G.}~\bibnamefont{Levy}}, \bibinfo {author}
  {\bibfnamefont{N.~P.}\ \bibnamefont{Butch}}, \bibinfo {author}
  {\bibfnamefont{P.}~\bibnamefont{Syers}}, \bibinfo {author}
  {\bibfnamefont{X.~F.}\ \bibnamefont{Wang}}, \bibinfo {author}
  {\bibfnamefont{J.}~\bibnamefont{Paglione}}, \bibinfo {author}
  {\bibfnamefont{G.~A.}\ \bibnamefont{Sawatzky}}, \bibinfo {author}
  {\bibfnamefont{I.~S.}\ \bibnamefont{Elfimov}},\ and\ \bibinfo {author}
  {\bibfnamefont{A.}~\bibnamefont{Damascelli}},\ }%
  \bibfield{journal}{%
  \bibinfo {journal} {Phys. Rev. Lett.}\ }%
  \textbf{\bibinfo {volume} {111}},\ \bibinfo {pages} {216402} (\bibinfo {year}
  {2013})%
  \bibAnnoteFile{NoStop}{sawatzky_smb6}%
\bibitem{smb6_prx_arpes}%
  \BibitemOpen
  \bibfield{author}{%
  \bibinfo {author} {\bibfnamefont{E.}~\bibnamefont{Frantzeskakis}}, \bibinfo
  {author} {\bibfnamefont{N.}~\bibnamefont{de~Jong}}, \bibinfo {author}
  {\bibfnamefont{B.}~\bibnamefont{Zwartsenberg}}, \bibinfo {author}
  {\bibfnamefont{Y.~K.}\ \bibnamefont{Huang}}, \bibinfo {author}
  {\bibfnamefont{Y.}~\bibnamefont{Pan}}, \bibinfo {author}
  {\bibfnamefont{X.}~\bibnamefont{Zhang}}, \bibinfo {author}
  {\bibfnamefont{J.~X.}\ \bibnamefont{Zhang}}, \bibinfo {author}
  {\bibfnamefont{F.~X.}\ \bibnamefont{Zhang}}, \bibinfo {author}
  {\bibfnamefont{L.~H.}\ \bibnamefont{Bao}}, \bibinfo {author}
  {\bibfnamefont{O.}~\bibnamefont{Tegus}}, \bibinfo {author}
  {\bibfnamefont{A.}~\bibnamefont{Varykhalov}}, \bibinfo {author}
  {\bibfnamefont{A.}~\bibnamefont{de~Visser}},\ and\ \bibinfo {author}
  {\bibfnamefont{M.~S.}\ \bibnamefont{Golden}},\ }%
  \bibfield{journal}{%
  \bibinfo {journal} {Phys. Rev. X}\ }%
  \textbf{\bibinfo {volume} {3}},\ \bibinfo {pages} {041024} (\bibinfo {year}
  {2013})%
  \bibAnnoteFile{NoStop}{smb6_prx_arpes}%
\bibitem{smb6_trivial}%
  \BibitemOpen
  \bibfield{author}{%
  \bibinfo {author} {\bibfnamefont{P.}~\bibnamefont{Hlawenka}}, \bibinfo
  {author} {\bibfnamefont{K.}~\bibnamefont{Siemensmeyer}}, \bibinfo {author}
  {\bibfnamefont{E.}~\bibnamefont{Weschke}}, \bibinfo {author}
  {\bibfnamefont{A.}~\bibnamefont{Varykhalova}}, \bibinfo {author}
  {\bibfnamefont{J.}~\bibnamefont{S\'{a}nchez-Barriga}}, \bibinfo {author}
  {\bibfnamefont{N.~Y.}\ \bibnamefont{Shitsevalova}}, \bibinfo {author}
  {\bibfnamefont{A.~V.}\ \bibnamefont{Dukhnenko}}, \bibinfo {author}
  {\bibfnamefont{V.~B.}\ \bibnamefont{Filipov}}, \bibinfo {author}
  {\bibfnamefont{S.}~\bibnamefont{Gabáni}}, \bibinfo {author}
  {\bibfnamefont{K.}~\bibnamefont{Flachbart}}, \bibinfo {author}
  {\bibfnamefont{O.}~\bibnamefont{Rader}},\ and\ \bibinfo {author}
  {\bibfnamefont{E.~D.~L.}\ \bibnamefont{Rienks}},\ }%
  \bibinfo {journal} {preprint arXiv:1502.01542}%
  \bibAnnoteFile{NoStop}{smb6_trivial}%
\bibitem{pnas_smb6_stm}%
  \BibitemOpen
\bibfield{journal}{%
    }%
  \bibfield{author}{%
  \bibinfo {author} {\bibfnamefont{S.}~\bibnamefont{R\"o{\ss}ler}}, \bibinfo
  {author} {\bibfnamefont{T.-H.}\ \bibnamefont{Jang}}, \bibinfo {author}
  {\bibfnamefont{D.-J.}\ \bibnamefont{Kim}}, \bibinfo {author}
  {\bibfnamefont{L.~H.}\ \bibnamefont{Tjeng}}, \bibinfo {author}
  {\bibfnamefont{Z.}~\bibnamefont{Fisk}}, \bibinfo {author}
  {\bibfnamefont{F.}~\bibnamefont{Steglich}},\ and\ \bibinfo {author}
  {\bibfnamefont{S.}~\bibnamefont{Wirth}},\ }%
  \bibfield{journal}{%
  \bibinfo {journal} {Proc. Nat. Acad. Sci.}\ }%
  \textbf{\bibinfo {volume} {111}},\ \bibinfo {pages} {4798} (\bibinfo {year}
  {2014})%
  \bibAnnoteFile{NoStop}{pnas_smb6_stm}%
\bibitem{hoffman_smb6}%
  \BibitemOpen
  \bibfield{author}{%
  \bibinfo {author} {\bibfnamefont{M.~M.}\ \bibnamefont{Yee}}, \bibinfo
  {author} {\bibfnamefont{Y.}~\bibnamefont{He}}, \bibinfo {author}
  {\bibfnamefont{A.}~\bibnamefont{Soumyanarayanan}}, \bibinfo {author}
  {\bibfnamefont{D.-J.}\ \bibnamefont{Kim}}, \bibinfo {author}
  {\bibfnamefont{Z.}~\bibnamefont{Fisk}},\ and\ \bibinfo {author}
  {\bibfnamefont{J.~E.}\ \bibnamefont{Hoffman}},\ }%
  \bibinfo {journal} {preprint arXiv:1308.1085}%
  \bibAnnoteFile{NoStop}{hoffman_smb6}%
\bibitem{pbsnse_tci}%
  \BibitemOpen
\bibfield{journal}{%
    }%
  \bibfield{author}{%
  \bibinfo {author} {\bibfnamefont{P.}~\bibnamefont{Dziawa}}, \bibinfo {author}
  {\bibfnamefont{B.~J.}\ \bibnamefont{Kowalski}}, \bibinfo {author}
  {\bibfnamefont{K.}~\bibnamefont{Dybko}}, \bibinfo {author}
  {\bibfnamefont{R.}~\bibnamefont{Buczko}}, \bibinfo {author}
  {\bibfnamefont{A.}~\bibnamefont{Szczerbakow}}, \bibinfo {author}
  {\bibfnamefont{M.}~\bibnamefont{Szot}}, \bibinfo {author}
  {\bibfnamefont{E.}~\bibnamefont{Lusakowska}}, \bibinfo {author}
  {\bibfnamefont{T.}~\bibnamefont{Balasubramanian}}, \bibinfo {author}
  {\bibfnamefont{B.~M.}\ \bibnamefont{Wojek}}, \bibinfo {author}
  {\bibfnamefont{M.~H.}\ \bibnamefont{Berntsen}}, \bibinfo {author}
  {\bibfnamefont{O.}~\bibnamefont{Tjernberg}},\ and\ \bibinfo {author}
  {\bibfnamefont{T.}~\bibnamefont{Story}},\ }%
  \bibfield{journal}{%
  \bibinfo {journal} {Nature Mat.}\ }%
  \textbf{\bibinfo {volume} {11}},\ \bibinfo {pages} {1023} (\bibinfo {year}
  {2012})%
  \bibAnnoteFile{NoStop}{pbsnse_tci}%
\bibitem{pbsnte_tci}%
  \BibitemOpen
  \bibfield{author}{%
  \bibinfo {author} {\bibfnamefont{S.-Y.}\ \bibnamefont{Xu}}, \bibinfo {author}
  {\bibfnamefont{C.}~\bibnamefont{Liu}}, \bibinfo {author}
  {\bibfnamefont{N.}~\bibnamefont{Alidoust}}, \bibinfo {author}
  {\bibfnamefont{M.}~\bibnamefont{Neupane}}, \bibinfo {author}
  {\bibfnamefont{D.}~\bibnamefont{Qian}}, \bibinfo {author}
  {\bibfnamefont{I.}~\bibnamefont{Belopolski}}, \bibinfo {author}
  {\bibfnamefont{J.}~\bibnamefont{Denlinger}}, \bibinfo {author}
  {\bibfnamefont{Y.}~\bibnamefont{Wang}}, \bibinfo {author}
  {\bibfnamefont{H.}~\bibnamefont{Lin}}, \bibinfo {author}
  {\bibfnamefont{L.}~\bibnamefont{Wray}}, \bibinfo {author}
  {\bibfnamefont{G.}~\bibnamefont{Landolt}}, \bibinfo {author}
  {\bibfnamefont{B.}~\bibnamefont{Slomski}}, \bibinfo {author}
  {\bibfnamefont{J.}~\bibnamefont{Dil}}, \bibinfo {author}
  {\bibfnamefont{A.}~\bibnamefont{Marcinkova}}, \bibinfo {author}
  {\bibfnamefont{E.}~\bibnamefont{Morosan}}, \bibinfo {author}
  {\bibfnamefont{Q.}~\bibnamefont{Gibson}}, \bibinfo {author}
  {\bibfnamefont{R.}~\bibnamefont{Sankar}}, \bibinfo {author}
  {\bibfnamefont{F.}~\bibnamefont{Chou}}, \bibinfo {author}
  {\bibfnamefont{R.}~\bibnamefont{Cava}}, \bibinfo {author}
  {\bibfnamefont{A.}~\bibnamefont{Bansil}},\ and\ \bibinfo {author}
  {\bibfnamefont{M.}~\bibnamefont{Hasan}},\ }%
  \bibfield{journal}{%
  \bibinfo {journal} {Nature Comm.}\ }%
  \textbf{\bibinfo {volume} {3}},\ \bibinfo {pages} {1192} (\bibinfo {year}
  {2012})%
  \bibAnnoteFile{NoStop}{pbsnte_tci}%
\bibitem{snte_tci}%
  \BibitemOpen
  \bibfield{author}{%
  \bibinfo {author} {\bibfnamefont{T.}~\bibnamefont{Hsieh}}, \bibinfo {author}
  {\bibfnamefont{H.}~\bibnamefont{Lin}}, \bibinfo {author}
  {\bibfnamefont{J.}~\bibnamefont{Liu}}, \bibinfo {author}
  {\bibfnamefont{W.}~\bibnamefont{Duan}}, \bibinfo {author}
  {\bibfnamefont{A.}~\bibnamefont{Bansil}},\ and\ \bibinfo {author}
  {\bibfnamefont{L.}~\bibnamefont{Fu}},\ }%
  \bibfield{journal}{%
  \bibinfo {journal} {Nature Comm.}\ }%
  \textbf{\bibinfo {volume} {3}},\ \bibinfo {pages} {982} (\bibinfo {year}
  {2012})%
  \bibAnnoteFile{NoStop}{snte_tci}%
\bibitem{smb6_stm_china}%
  \BibitemOpen
  \bibfield{author}{%
  \bibinfo {author} {\bibfnamefont{W.}~\bibnamefont{Ruan}}, \bibinfo {author}
  {\bibfnamefont{C.}~\bibnamefont{Ye}}, \bibinfo {author}
  {\bibfnamefont{M.}~\bibnamefont{Guo}}, \bibinfo {author}
  {\bibfnamefont{F.}~\bibnamefont{Chen}}, \bibinfo {author}
  {\bibfnamefont{X.}~\bibnamefont{Chen}}, \bibinfo {author}
  {\bibfnamefont{G.-M.}\ \bibnamefont{Zhang}},\ and\ \bibinfo {author}
  {\bibfnamefont{Y.}~\bibnamefont{Wang}},\ }%
  \bibfield{journal}{%
  \bibinfo {journal} {Phys. Rev. Lett.}\ }%
  \textbf{\bibinfo {volume} {112}},\ \bibinfo {pages} {136401} (\bibinfo {year}
  {2014})%
  \bibAnnoteFile{NoStop}{smb6_stm_china}%
\bibitem{smb6_korea2}%
  \BibitemOpen
  \bibfield{author}{%
  \bibinfo {author} {\bibfnamefont{J.}~\bibnamefont{Kim}}, \bibinfo {author}
  {\bibfnamefont{K.}~\bibnamefont{Kim}}, \bibinfo {author}
  {\bibfnamefont{C.-J.}\ \bibnamefont{Kang}}, \bibinfo {author}
  {\bibfnamefont{S.}~\bibnamefont{Kim}}, \bibinfo {author}
  {\bibfnamefont{H.~C.}\ \bibnamefont{Choi}}, \bibinfo {author}
  {\bibfnamefont{J.-S.}\ \bibnamefont{Kang}}, \bibinfo {author}
  {\bibfnamefont{J.~D.}\ \bibnamefont{Denlinger}},\ and\ \bibinfo {author}
  {\bibfnamefont{B.~I.}\ \bibnamefont{Min}},\ }%
  \bibfield{journal}{%
  \bibinfo {journal} {Phys. Rev. B}\ }%
  \textbf{\bibinfo {volume} {90}},\ \bibinfo {pages} {075131} (\bibinfo {year}
  {2014})%
  \bibAnnoteFile{NoStop}{smb6_korea2}%
\bibitem{pub6}%
  \BibitemOpen
  \bibfield{author}{%
  \bibinfo {author} {\bibfnamefont{X.}~\bibnamefont{Deng}}, \bibinfo {author}
  {\bibfnamefont{K.}~\bibnamefont{Haule}},\ and\ \bibinfo {author}
  {\bibfnamefont{G.}~\bibnamefont{Kotliar}},\ }%
  \bibfield{journal}{%
  \bibinfo {journal} {Phys. Rev. Lett.}\ }%
  \textbf{\bibinfo {volume} {111}},\ \bibinfo {pages} {176404} (\bibinfo {year}
  {2013})%
  \bibAnnoteFile{NoStop}{pub6}%
\bibitem{prbr_io_smb6}%
  \BibitemOpen
  \bibfield{author}{%
  \bibinfo {author} {\bibfnamefont{P.~P.}\ \bibnamefont{Baruselli}}\ and\
  \bibinfo {author} {\bibfnamefont{M.}~\bibnamefont{Vojta}},\ }%
  \bibfield{journal}{%
  \bibinfo {journal} {Phys. Rev. B}\ }%
  \textbf{\bibinfo {volume} {90}},\ \bibinfo {pages} {201106} (\bibinfo {year}
  {2014})%
  \bibAnnoteFile{NoStop}{prbr_io_smb6}%
\bibitem{smb6_dzero_pert}%
  \BibitemOpen
  \bibfield{author}{%
  \bibinfo {author} {\bibfnamefont{B.}~\bibnamefont{Roy}}, \bibinfo {author}
  {\bibfnamefont{J.~D.}\ \bibnamefont{Sau}}, \bibinfo {author}
  {\bibfnamefont{M.}~\bibnamefont{Dzero}},\ and\ \bibinfo {author}
  {\bibfnamefont{V.}~\bibnamefont{Galitski}},\ }%
  \bibfield{journal}{%
  \bibinfo {journal} {Phys. Rev. B}\ }%
  \textbf{\bibinfo {volume} {90}},\ \bibinfo {pages} {155314} (\bibinfo {year}
  {2014})%
  \bibAnnoteFile{NoStop}{smb6_dzero_pert}%
\bibitem{fehske}%
  \BibitemOpen
  \bibfield{author}{%
  \bibinfo {author} {\bibfnamefont{G.}~\bibnamefont{Schubert}}, \bibinfo
  {author} {\bibfnamefont{H.}~\bibnamefont{Fehske}}, \bibinfo {author}
  {\bibfnamefont{L.}~\bibnamefont{Fritz}},\ and\ \bibinfo {author}
  {\bibfnamefont{M.}~\bibnamefont{Vojta}},\ }%
  \bibfield{journal}{%
  \bibinfo {journal} {Phys. Rev. B}\ }%
  \textbf{\bibinfo {volume} {85}},\ \bibinfo {pages} {201105(R)} (\bibinfo
  {year} {2012})%
  \bibAnnoteFile{NoStop}{fehske}%
\bibitem{smb6_iondamaged}%
  \BibitemOpen
  \bibfield{author}{%
  \bibinfo {author} {\bibfnamefont{N.}~\bibnamefont{Wakeham}}, \bibinfo
  {author} {\bibfnamefont{Y.~Q.}\ \bibnamefont{Wang}}, \bibinfo {author}
  {\bibfnamefont{Z.}~\bibnamefont{Fisk}}, \bibinfo {author}
  {\bibfnamefont{F.}~\bibnamefont{Ronning}},\ and\ \bibinfo {author}
  {\bibfnamefont{J.~D.}\ \bibnamefont{Thompson}},\ }%
  \bibfield{journal}{%
  \bibinfo {journal} {Phys. Rev. B}\ }%
  \textbf{\bibinfo {volume} {91}},\ \bibinfo {pages} {085107} (\bibinfo {year}
  {2015})%
  \bibAnnoteFile{NoStop}{smb6_iondamaged}%
\bibitem{hewson}%
  \BibitemOpen
  \bibfield{author}{%
  \bibinfo {author} {\bibfnamefont{A.~C.}\ \bibnamefont{Hewson}},\ }%
  \emph{\bibinfo {title} {The Kondo Problem to Heavy Fermions}}\ (\bibinfo
  {publisher} {Cambridge University Press, Cambridge},\ \bibinfo {year}
  {1993})%
  \bibAnnoteFile{NoStop}{hewson}%
\bibitem{renorm_foot}%
  \BibitemOpen
  \bibinfo {note} {Interaction-induced renormalizations typically also reduce
  the quasiparticle weight in both ARPES and QPI. This is not taken into
  account in our modelling.}%
  \bibAnnoteFile{Stop}{renorm_foot}%
\bibitem{prb_io_tki}%
  \BibitemOpen
  \bibfield{author}{%
  \bibinfo {author} {\bibfnamefont{P.~P.}\ \bibnamefont{Baruselli}}\ and\
  \bibinfo {author} {\bibfnamefont{M.}~\bibnamefont{Vojta}},\ }%
  \bibfield{journal}{%
  \bibinfo {journal} {Phys. Rev. B}\ }%
  \textbf{\bibinfo {volume} {89}},\ \bibinfo {pages} {205105} (\bibinfo {year}
  {2014})%
  \bibAnnoteFile{NoStop}{prb_io_tki}%
\bibitem{kondo_breakdown}%
  \BibitemOpen
  \bibfield{author}{%
  \bibinfo {author} {\bibfnamefont{V.}~\bibnamefont{Alexandrov}}, \bibinfo
  {author} {\bibfnamefont{P.}~\bibnamefont{Coleman}},\ and\ \bibinfo {author}
  {\bibfnamefont{O.}~\bibnamefont{Erten}},\ }%
  \bibfield{journal}{%
  \bibinfo {journal} {Phys. Rev. Lett.}\ }%
  \textbf{\bibinfo {volume} {114}},\ \bibinfo {pages} {177202} (\bibinfo {year}
  {2015})%
  \bibAnnoteFile{NoStop}{kondo_breakdown}%
\bibitem{smb6_broholm_exciton}%
  \BibitemOpen
  \bibfield{author}{%
  \bibinfo {author} {\bibfnamefont{W.~T.}\ \bibnamefont{Fuhrman}}, \bibinfo
  {author} {\bibfnamefont{J.}~\bibnamefont{Leiner}}, \bibinfo {author}
  {\bibfnamefont{P.}~\bibnamefont{Nikoli\ifmmode~\acute{c}\else \'{c}\fi{}}},
  \bibinfo {author} {\bibfnamefont{G.~E.}\ \bibnamefont{Granroth}}, \bibinfo
  {author} {\bibfnamefont{M.~B.}\ \bibnamefont{Stone}}, \bibinfo {author}
  {\bibfnamefont{M.~D.}\ \bibnamefont{Lumsden}}, \bibinfo {author}
  {\bibfnamefont{L.}~\bibnamefont{DeBeer-Schmitt}}, \bibinfo {author}
  {\bibfnamefont{P.~A.}\ \bibnamefont{Alekseev}}, \bibinfo {author}
  {\bibfnamefont{J.-M.}\ \bibnamefont{Mignot}}, \bibinfo {author}
  {\bibfnamefont{S.~M.}\ \bibnamefont{Koohpayeh}}, \bibinfo {author}
  {\bibfnamefont{P.}~\bibnamefont{Cottingham}}, \bibinfo {author}
  {\bibfnamefont{W.~A.}\ \bibnamefont{Phelan}}, \bibinfo {author}
  {\bibfnamefont{L.}~\bibnamefont{Schoop}}, \bibinfo {author}
  {\bibfnamefont{T.~M.}\ \bibnamefont{McQueen}},\ and\ \bibinfo {author}
  {\bibfnamefont{C.}~\bibnamefont{Broholm}},\ }%
  \bibfield{journal}{%
  \bibinfo {journal} {Phys. Rev. Lett.}\ }%
  \textbf{\bibinfo {volume} {114}},\ \bibinfo {pages} {036401} (\bibinfo
  {month} {Jan}\ \bibinfo {year} {2015})%
  \bibAnnoteFile{NoStop}{smb6_broholm_exciton}%
\bibitem{smb6_denlinger_t}%
  \BibitemOpen
  \bibfield{author}{%
  \bibinfo {author} {\bibfnamefont{J.~D.}\ \bibnamefont{Denlinger}}, \bibinfo
  {author} {\bibfnamefont{J.~W.}\ \bibnamefont{Allen}}, \bibinfo {author}
  {\bibfnamefont{J.-S.}\ \bibnamefont{Kang}}, \bibinfo {author}
  {\bibfnamefont{K.}~\bibnamefont{Sun}}, \bibinfo {author}
  {\bibfnamefont{J.-W.}\ \bibnamefont{Kim}}, \bibinfo {author}
  {\bibfnamefont{J.}~\bibnamefont{Shim}}, \bibinfo {author}
  {\bibfnamefont{B.~I.}\ \bibnamefont{Min}}, \bibinfo {author}
  {\bibfnamefont{D.~J.}\ \bibnamefont{Kim}},\ and\ \bibinfo {author}
  {\bibfnamefont{F.}~\bibnamefont{Fisk}},\ }%
  \bibinfo {journal} {preprint arXiv:1312.6637}%
  \bibAnnoteFile{NoStop}{smb6_denlinger_t}%
\bibitem{snte_tci_arpes}%
  \BibitemOpen
\bibfield{journal}{%
    }%
  \bibfield{author}{%
  \bibinfo {author} {\bibfnamefont{Y.}~\bibnamefont{Y.~Tanaka}}, \bibinfo
  {author} {\bibfnamefont{Z.}~\bibnamefont{Ren}}, \bibinfo {author}
  {\bibfnamefont{T.}~\bibnamefont{Sato}}, \bibinfo {author}
  {\bibfnamefont{K.}~\bibnamefont{Nakayama}}, \bibinfo {author}
  {\bibfnamefont{S.}~\bibnamefont{Souma}}, \bibinfo {author}
  {\bibfnamefont{T.}~\bibnamefont{Takahashi}}, \bibinfo {author}
  {\bibfnamefont{K.}~\bibnamefont{Segawa}},\ and\ \bibinfo {author}
  {\bibfnamefont{Y.}~\bibnamefont{Ando}},\ }%
  \bibfield{journal}{%
  \bibinfo {journal} {Nat. Phys.}\ }%
  \textbf{\bibinfo {volume} {8}},\ \bibinfo {pages} {800} (\bibinfo {year}
  {2012})%
  \bibAnnoteFile{NoStop}{snte_tci_arpes}%
\bibitem{kane_bisb}%
  \BibitemOpen
  \bibfield{author}{%
  \bibinfo {author} {\bibfnamefont{J.~C.~Y.}\ \bibnamefont{Teo}}, \bibinfo
  {author} {\bibfnamefont{L.}~\bibnamefont{Fu}},\ and\ \bibinfo {author}
  {\bibfnamefont{C.~L.}\ \bibnamefont{Kane}},\ }%
  \bibfield{journal}{%
  \bibinfo {journal} {Phys. Rev. B}\ }%
  \textbf{\bibinfo {volume} {78}},\ \bibinfo {pages} {045426} (\bibinfo {year}
  {2008})%
  \bibAnnoteFile{NoStop}{kane_bisb}%
\bibitem{sigrist_tki_prb}%
  \BibitemOpen
  \bibfield{author}{%
  \bibinfo {author} {\bibfnamefont{M.}~\bibnamefont{Legner}}, \bibinfo {author}
  {\bibfnamefont{A.}~\bibnamefont{R\"uegg}},\ and\ \bibinfo {author}
  {\bibfnamefont{M.}~\bibnamefont{Sigrist}},\ }%
  \bibfield{journal}{%
  \bibinfo {journal} {Phys. Rev. B}\ }%
  \textbf{\bibinfo {volume} {89}},\ \bibinfo {pages} {085110} (\bibinfo {year}
  {2014})%
  \bibAnnoteFile{NoStop}{sigrist_tki_prb}%
\bibitem{xue}%
  \BibitemOpen
  \bibfield{author}{%
  \bibinfo {author} {\bibfnamefont{T.}~\bibnamefont{Zhang}}, \bibinfo {author}
  {\bibfnamefont{P.}~\bibnamefont{Cheng}}, \bibinfo {author}
  {\bibfnamefont{X.}~\bibnamefont{Chen}}, \bibinfo {author}
  {\bibfnamefont{J.-F.}\ \bibnamefont{Jia}}, \bibinfo {author}
  {\bibfnamefont{X.}~\bibnamefont{Ma}}, \bibinfo {author}
  {\bibfnamefont{K.}~\bibnamefont{He}}, \bibinfo {author}
  {\bibfnamefont{L.}~\bibnamefont{Wang}}, \bibinfo {author}
  {\bibfnamefont{H.}~\bibnamefont{Zhang}}, \bibinfo {author}
  {\bibfnamefont{X.}~\bibnamefont{Dai}}, \bibinfo {author}
  {\bibfnamefont{Z.}~\bibnamefont{Fang}}, \bibinfo {author}
  {\bibfnamefont{X.}~\bibnamefont{Xie}},\ and\ \bibinfo {author}
  {\bibfnamefont{Q.-K.}\ \bibnamefont{Xue}},\ }%
  \bibfield{journal}{%
  \bibinfo {journal} {Phys. Rev. Lett.}\ }%
  \textbf{\bibinfo {volume} {103}},\ \bibinfo {pages} {266803} (\bibinfo {year}
  {2009})%
  \bibAnnoteFile{NoStop}{xue}%
\bibitem{yazdani}%
  \BibitemOpen
  \bibfield{author}{%
  \bibinfo {author} {\bibfnamefont{P.}~\bibnamefont{Roushan}}, \bibinfo
  {author} {\bibfnamefont{J.}~\bibnamefont{Seo}}, \bibinfo {author}
  {\bibfnamefont{C.~V.}\ \bibnamefont{Parker}}, \bibinfo {author}
  {\bibfnamefont{Y.~S.}\ \bibnamefont{Hor}}, \bibinfo {author}
  {\bibfnamefont{D.}~\bibnamefont{Hsieh}}, \bibinfo {author}
  {\bibfnamefont{D.}~\bibnamefont{Qian}}, \bibinfo {author}
  {\bibfnamefont{A.}~\bibnamefont{Richardella}},\ and\ \bibinfo {author}
  {\bibfnamefont{M.~Z.}\ \bibnamefont{Hasan}},\ }%
  \bibfield{journal}{%
  \bibinfo {journal} {Nature}\ }%
  \textbf{\bibinfo {volume} {460}},\ \bibinfo {pages} {1106} (\bibinfo {year}
  {2009})%
  \bibAnnoteFile{NoStop}{yazdani}%
\bibitem{guo_franz_sti}%
  \BibitemOpen
  \bibfield{author}{%
  \bibinfo {author} {\bibfnamefont{H.-M.}\ \bibnamefont{Guo}}\ and\ \bibinfo
  {author} {\bibfnamefont{M.}~\bibnamefont{Franz}},\ }%
  \bibfield{journal}{%
  \bibinfo {journal} {Phys. Rev. B}\ }%
  \textbf{\bibinfo {volume} {81}},\ \bibinfo {pages} {041102} (\bibinfo {year}
  {2010})%
  \bibAnnoteFile{NoStop}{guo_franz_sti}%
\bibitem{tci_qpi_theory}%
  \BibitemOpen
  \bibfield{author}{%
  \bibinfo {author} {\bibfnamefont{C.}~\bibnamefont{Fang}}, \bibinfo {author}
  {\bibfnamefont{M.~J.}\ \bibnamefont{Gilbert}}, \bibinfo {author}
  {\bibfnamefont{S.-Y.}\ \bibnamefont{Xu}}, \bibinfo {author}
  {\bibfnamefont{B.~A.}\ \bibnamefont{Bernevig}},\ and\ \bibinfo {author}
  {\bibfnamefont{M.~Z.}\ \bibnamefont{Hasan}},\ }%
  \bibfield{journal}{%
  \bibinfo {journal} {Phys. Rev. B}\ }%
  \textbf{\bibinfo {volume} {88}},\ \bibinfo {pages} {125141} (\bibinfo {year}
  {2013})%
  \bibAnnoteFile{NoStop}{tci_qpi_theory}%
\bibitem{liu_qpi_spa}%
  \BibitemOpen
  \bibfield{author}{%
  \bibinfo {author} {\bibfnamefont{Q.}~\bibnamefont{Liu}}, \bibinfo {author}
  {\bibfnamefont{X.-L.}\ \bibnamefont{Qi}},\ and\ \bibinfo {author}
  {\bibfnamefont{S.-C.}\ \bibnamefont{Zhang}},\ }%
  \bibfield{journal}{%
  \bibinfo {journal} {Phys. Rev. B}\ }%
  \textbf{\bibinfo {volume} {85}},\ \bibinfo {pages} {125314} (\bibinfo {year}
  {2012})%
  \bibAnnoteFile{NoStop}{liu_qpi_spa}%
\bibitem{hoffman_private_comm}%
  \BibitemOpen
  \bibinfo {note} {J. E. Hoffman, private communication}%
  \bibAnnoteFile{NoStop}{hoffman_private_comm}%
\bibitem{coleman09}%
  \BibitemOpen
  \bibfield{author}{%
  \bibinfo {author} {\bibfnamefont{M.}~\bibnamefont{Maltseva},
  \bibfnamefont{M.~Dzero}}\ and\ \bibinfo {author}
  {\bibfnamefont{P.}~\bibnamefont{Coleman}},\ }%
  \bibfield{journal}{%
  \bibinfo {journal} {Phys. Rev. Lett.}\ }%
  \textbf{\bibinfo {volume} {103}},\ \bibinfo {pages} {206402} (\bibinfo {year}
  {2009})%
  \bibAnnoteFile{NoStop}{coleman09}%
\bibitem{zabol09}%
  \BibitemOpen
  \bibfield{author}{%
  \bibinfo {author} {\bibfnamefont{V.~B.}\ \bibnamefont{Zabolotnyy}}, \bibinfo
  {author} {\bibfnamefont{A.~A.}\ \bibnamefont{Kordyuk}}, \bibinfo {author}
  {\bibfnamefont{D.~S.}\ \bibnamefont{Inosov}}, \bibinfo {author}
  {\bibfnamefont{D.~V.}\ \bibnamefont{Evtushinsky}}, \bibinfo {author}
  {\bibfnamefont{R.}~\bibnamefont{Schuster}}, \bibinfo {author}
  {\bibfnamefont{B.}~\bibnamefont{B\"uchner}}, \bibinfo {author}
  {\bibfnamefont{N.}~\bibnamefont{Wizent}}, \bibinfo {author}
  {\bibfnamefont{G.}~\bibnamefont{Behr}}, \bibinfo {author}
  {\bibfnamefont{S.}~\bibnamefont{Pyon}}, \bibinfo {author}
  {\bibfnamefont{H.}~\bibnamefont{Takagi}}, \bibinfo {author}
  {\bibfnamefont{R.}~\bibnamefont{Follath}},\ and\ \bibinfo {author}
  {\bibfnamefont{S.~V.}\ \bibnamefont{Borisenko}},\ }%
  \bibfield{journal}{%
  \bibinfo {journal} {EPL}\ }%
  \textbf{\bibinfo {volume} {86}},\ \bibinfo {pages} {47005} (\bibinfo {year}
  {2009})%
  \bibAnnoteFile{NoStop}{zabol09}%
\end{thebibliography}%

\end{document}